\documentclass[aps,prm,reprint,superscriptaddress,twocolumn]{revtex4}

\usepackage{graphicx}
\usepackage{enumerate}
\usepackage{hyperref}
\usepackage{xcolor}
\usepackage{lscape}
\usepackage{longtable}

\newcommand{\strb}{{\em Strukturbericht}}

\newcommand{\csdb}{http://www.aflow.org/prototype-encyclopedia}

\makeatletter \renewcommand\frontmatter@abstractwidth{\dimexpr\textwidth\relax} \makeatother

\begin{document}

\title{The Tin Pest Problem as a Test of Density Functionals Using
  High-Throughput Calculations}

\author{Michael J. Mehl}
\email{michael.mehl@duke.edu}
\affiliation{Center for Autonomous Materials Design, Duke University, Durham NC 27708}

\author{Mateo Ronquillo}
\affiliation{U. S. Nuclear Power School, Goose Creek, South Carolina 29445}

\author{David Hicks}
\affiliation{Center of Autonomous Materials Design, Duke University, Durham NC 27708}

\author{Marco Esters}
\affiliation{Center of Autonomous Materials Design, Duke University, Durham NC 27708}

\author{Corey Oses}

\affiliation{Center of Autonomous Materials Design, Duke University, Durham NC 27708}

\author{Rico Friedrich}
\affiliation{Center of Autonomous Materials Design, Duke University, Durham NC 27708}

\author{Andriy Smolyanyuk}
\affiliation{Center of Autonomous Materials Design, Duke University, Durham NC 27708}

\author{Eric Gossett}
\affiliation{Center of Autonomous Materials Design, Duke University, Durham NC 27708}

\author{Daniel Finkenstadt}
\affiliation{Physics Department, U.S. Naval Academy, Annapolis, Maryland 21402}

\author{Stefano Curtarolo}
\email{stefano@duke.edu}
\affiliation{Center of Autonomous Materials Design, Duke University, Durham NC 27708}

\date{\today}

\begin{abstract}
At ambient pressure tin transforms from its ground-state semi-metal
$\alpha$-Sn (diamond structure) phase to the compact metallic
$\beta$-Sn phase at 13$^\circ$C (286K).  There may be a further
transition to the simple hexagonal $\gamma$-Sn above 450K. These
relatively low transition temperatures are due to the small energy
differences between the structures, $\approx 20$\,meV/atom between
$\alpha$- and $\beta$-Sn.  This makes tin an exceptionally sensitive
test of the accuracy of density functionals and computational methods.
Here we use the high-throughput Automatic-FLOW (AFLOW) method to study
the energetics of tin in multiple structures using a variety of
density functionals.  We look at the successes and deficiencies of
each functional.  As no functional is completely satisfactory, we look
Hubbard U corrections and show that the Coulomb interaction can be
chosen to predict the correct phase transition temperature.  We also
discuss the necessity of testing high-throughput calculations for
convergence for systems with small energy differences.
\end{abstract}

\maketitle

\section{\label{sec:intro} Introduction}

Tin is one of the few elements which undergoes a near-room-temperature
(286K) first-order structural phase transition from semi-metallic
$\alpha$-Sn (gray tin, diamond, \strb{} designation $A4$, AFLOW
label\cite{Mehl_Comp_Mat_Sci_136_S1_2017} A\_cF8\_227\_a) to metallic
$\beta$-Sn (white tin, \strb{} designation $A5$, AFLOW label
A\_tI4\_141\_a).\cite{Cohen_Z_Phys_Chem_173_A_32_1935} This transition
takes place slowly and is accompanied by a large (20\%) reduction in
the volume.  On cooling the resulting expansion leads to the
``tin-pest'' problem when $\beta$-Sn is stored at low
temperatures.\cite{Cornelius_Microelec_Rel_79_175_2017} Consequences
of this have occurred in the real
world\cite{Fritzsche_Berichte_Deutsche_Chem_Gesell_1869,Plumbridge_JMS_Mater_Elec_18_307_2007}
and in legend.\cite{LeCouteur_Napoleons_Buttons_2004} Today, the main
concern about tin-pest is in lead-free
solders,\cite{Pfister_arxiv_1204_1443_2012} as they seem to be more
susceptible to the problem than the traditional tin-lead amalgam.

In addition to its well-known phases, tin may undergo a transition
from $\beta$-Sn to $\gamma$-Sn at 450K.\cite{Kubiak_JLCM_116_307_1986}
This phase is said to be a slight orthorhombic distortion of the
simple hexagonal lattice (\strb{} designation $A_{f}$, Pearson symbol
$hP1$, space group $P6/mmm \# 191$, AFLOW label A\_hP1\_191\_a) a
structure not observed at ambient pressure in any other element. It
can be stabilized in the hexagonal $\gamma$-Sn form by alloying with
cadmium, indium, lead and
mercury.\cite{Kane_Acta_Met_14_605_1966,Ivanov_J_Phys_F_17_1925_1987,Ivanov_Physica_B_174_79_1991}
Alloys such as In$_{0.45}$Bi$_{0.55}$ are also found in this
structure.\cite{Parthe_Gmelin_Handbook_1993}

The static lattice energy difference between $\alpha$- and $\beta$-Sn
has not been directly determined by experiment, but it is estimated to
be in the range
10-40\,meV/atom.\cite{Pavone_PRB_57_10421_1998,Houben_PRB_100_075408_2019}
This makes the prediction of the the $\alpha$-Sn $\rightarrow
\beta$-Sn transition difficult for density functional (DFT)
calculations, which may not achieve the required
accuracy.\cite{Pavone_PRB_57_10421_1998,Ihm_PRB_23_1576_1981,Na_J_Korean_Phys_Soc_54_494_2010,Legrain_JCP_143_204701_2015,Legrain_AIPAdvances_6_045116_2016}
In fact, as we shall see, different functionals will give different
predictions for the $\alpha \rightarrow \beta$ transition temperature,
some will find $\beta$-Sn as the ground state, and others will predict
the ground state of tin to be close-packed structure.  Tin then
becomes an ideal system for testing the accuracy of density
functionals, including supposedly more accurate generalized-gradient
(GGA) and meta-GGA functionals.

Since the tin phase transition is thermal, its prediction requires
accurate static-lattice energy and vibrational spectra for the
$\alpha$-, $\beta$-, and $\gamma$-Sn as well as static-lattice
energies for the close-packed phases.  This large number of
calculations, involving supercells of up to 256 atoms, is best handled
by high-throughput methods, but this introduces another source of
error: the basis-set and k-point mesh sizes set by default in these
programs might not be accurate enough to find the correct ordering of
phases.  Thorough testing of the predictive capability of different
functionals also requires testing of convergence criteria in the
programs that evaluate DFT energies.

In this paper we will test several DFTs, determining the static
lattice energy of tin at multiple volumes in a variety of crystal
structures, computing the corresponding phonon spectra for the
$\alpha$-, $\beta$-, and $\gamma$-Sn phases, and evaluating the free
energy for each structures as a function of temperature.  These
calculations are most easily accomplished via high-throughput
calculations and we use the AFLOW (Automatic FLOW)
platform\cite{Curtarolo_Comp_Mat_Sci_58_218_2012,Toher_AFLOW_Handbook_Mat_Model_1_2018,Oses_MRS_Bull_43_670_2018,curtarolo:art113,curtarolo:art67,curtarolo:art139},
with the Vienna {\em Ab initio} Simulation Package (VASP) to perform
the first-principles
calculations,\cite{Kresse_PRB_47_558_1993,Kresse_PRB_49_14251_1994,Kresse_Comp_Mat_Sci_6_15_1996,Kresse_PRB_54_11169_1996}
allowing AFLOW set up, run, and interpret the calculations using its
harmonic phonon module (APL)\cite{curtarolo:art65,curtarolo:art125}.

Our strategy is to use AFLOW and VASP to look at possible elemental
structures for tin using local, generalized-gradient, and meta-GGA
density functionals available in VASP.  We previously
showed\cite{Mehl_PRB_91_184100_2015} that different functionals can
predict quite different ground state structures.  Here a functional
must predict that $\alpha$-Sn is the ground state of tin and that the
$\beta$-Sn phase is close enough in energy to it that a
room-temperature thermal phase transition is possible.  If it is, we
then compute the phonon spectra of these phases as a function of
volume, use this to find the free energy as a function of temperature
within the quasi-harmonic approximation, and determine the
functional's prediction of the transition temperature.

Finally, we note that the default settings for high-throughput
calculations need to be carefully checked in situations where there
are small energy differences.  Here we show that different choices of
basis-sets and k-points meshes can give different results for the
lowest energy structure.

The paper is organized as follows: computational details are provided
in Section \ref{sec:computational}.  Section \ref{sec:structures}
describes the crystal structures investigated in this paper, including
the many ways the structures are referred to in the literature.
Section \ref{sec:dftresults} summarizes our results for the LDA, GGA,
and meta-GGA density functionals available in VASP.
Section~\ref{sec:thermal} considers the thermal properties of the
$\alpha$-, $\beta$-, and $\gamma$-Sn phases, including predictions of
phase transition temperatures.  In Section~\ref{sec:hubbard} we apply
a Hubbard correction to the LDA and PBE functionals to see if we can
improve our results.  In this we follow Legrain and
Manzhos\cite{Legrain_AIPAdvances_6_045116_2016}, using the VASP
implementation of
LDAU\cite{Bengone_PRB_62_16392_2000,Rohrbach_JPCM_15_979_2003} to
determine the changes made by applying a Hubbard $U$ to either the
$s$- or $p$-valence orbitals in tin, and show that with one adjustable
parameter we can correctly predict the 286K phase transition.
Section~\ref{sec:convergence} discusses the problems of convergence in
systems with small energy differences between phases.  Finally,
section~\ref{sec:conclude} summarizes our results.

\section{\label{sec:computational} Computational Details}

All calculations were performed using the high-throughput AFLOW
(Automatic
FLOW)\cite{Curtarolo_Comp_Mat_Sci_58_218_2012,Toher_AFLOW_Handbook_Mat_Model_1_2018,Oses_MRS_Bull_43_670_2018}
framework, using Vienna {\em Ab initio} Simulation Package (VASP)
version 5.4.4 to perform the first-principles
calculations,\cite{Kresse_PRB_47_558_1993,Kresse_PRB_49_14251_1994,Kresse_Comp_Mat_Sci_6_15_1996,Kresse_PRB_54_11169_1996}
using the Projector Augmented-Wave (PAW)
method\cite{Blochl_PRB_50_17953_1994,Kresse_PRB_59_1758_1999} with the
standard VASP POTCAR files.  Calculations with the
Perdew-Burke-Ernzerhof (PBE) functional\cite{Perdew_PRL_77_3865_1996}
used the VASP $s^2p^2$ ``Sn'' PBE POTCAR (dated 08Apr2002).  Local
Density Approximation (LDA) and other GGAs used the corresponding LDA
POTCAR (03Oct2001), with the appropriate choice of the GGA or METAGGA
tag in the INCAR file.  Meta-GGA functional calculations require
kinetic energy information only available in POTCARs available
starting with VASP 5.4, so for those we used the $s^2p^2d^{10}$ LDA
POTCAR (dated 09Feb1998).  In general we used the AFLOW defaults for
energy cutoffs and $\Gamma$-centered k-point meshes, with the
exceptions noted below.

Phonon spectra, vibrational free energy, and the thermal expansion of
the $\alpha$-, $\beta$-, and simple hexagonal $\gamma$-Sn structures
were done using the Automatic Phonon Library (APL) module in AFLOW.
The phonon spectra of both $\beta$- and $\gamma$-Sn are extremely
sensitive to the k-point mesh size, so we had to adjust the default
settings in APL to get good results while not over-burdening our
computational platforms.

\begin{itemize}
\item For $\alpha$-Sn we used the standard face-centered cubic lattice
  vectors.  APL calculations used a $5 \times 5 \times 5$ (250 atom)
  supercell and a $2 \times 2 \times 2$ $\Gamma$-centered k-point
  mesh, yielding 6 k-points in the irreducible part of the supercell's
  Brillouin zone.
\item $\beta$-Sn has a very small value for $c/a$ ($\approx 0.54$), which
  required some special handling.  We used the primitive vectors
  \begin{eqnarray}
    \vec{a}_1 & = & a \, \hat{x} \nonumber \\
    \vec{a}_2 & = & a \, \hat{y} ~ , ~ \mbox{and} \nonumber \\
    \vec{a}_3 & = & \frac12 a \, \hat{x} + \frac12 a \, \hat{y} +
    \frac12 c \, \hat{z}
    \label{equ:betaprim}
  \end{eqnarray}
  for the body-centered-tetragonal primitive vectors.  APL
  calculations used a $4 \times 4 \times 8$ (256 atom) supercell,
  corresponding to a nearly-cubic $4a \times 4a \times 4c$ tetragonal
  cell.  A $2 \times 2 \times 2$ $\Gamma$-centered k-point mesh was
  chosen, yielding 8 k-points in the irreducible part of the
  supercell's Brillouin zone.
\item We used a $6 \times 6 \times 6$ (216 atom) supercell for the APL
  calculations for $\gamma$-Sn.  The phonon spectra for this structure
  was extremely sensitive to the choice of k-point mesh, and we
  finally settled on a $6 \times 6 \times 4$ $\Gamma$-centered mesh
  with 47 k-points in the irreducible part of the supercell's
  Brillouin zone.
\end{itemize}

For a given primitive cell volume ($V$) we calculated the free energy
as a function of temperature for $\alpha$-, $\beta$- and $\gamma$-Sn
by starting with the energy determined from the static lattice,
$U(V)$.  The AFLOW APL module then determine the phonon spectrum
at this volume as well as the temperature-dependent vibrational energy
$U_{ph}(V,T)$ and entropy $S_{ph}(V,T)$.  The free energy of the
system at temperature $T$ is given by:
\begin{equation}
  F(V,T) = U(V) + U_{ph}(V,T) - T \, S_{ph}(V,T) ~ .
  \label{equ:freeeng}
\end{equation}
$U(V)$, $U_{ph}(V,T)$ and $S_{ph}(V,T)$ were computed for a number
of volumes $V_{n}$ around the equilibrium point.  At a given
temperature the minimum free-energy $F(T)$ was determined by fitting
the points $F(V_{n},T)$ to the third-order Birch equation of
state\cite{Birch_JGR_83_1257_1978} in the
form\cite{Mehl_PRB_47_2493_1993}
\begin{eqnarray}
  \label{equ:birch}
  F(V) & = & F_{0}(T) + \frac98 K_{0}(T) V_{0}(T)
  \left[\left(\frac{V_{0}(T)}{V}\right)^{2/3} - 1\right]^2  \nonumber \\
  & + & \frac{9}{16} K_{0}(T) V_{0}(T) (K_{0}(T)' - 4)
  \left[\left(\frac{V_{0}(T)}{V}\right)^{2/3} - 1\right]^3,
\end{eqnarray}
where $V_{0}$ is the equilibrium volume, $F_{0}$ the equilibrium free
energy, $K_{0}$ the equilibrium bulk modulus, and $K_{0}'$ the
pressure derivative of the bulk modulus at equilibrium.  The value
$F_{0}(T)$ in this calculation was taken to be the true free energy at
the given temperature, and the associated volume, $V_{0}(T)$, can be
used to compute the thermal expansion of the system and compare it
with experiment.

While we only need the primitive cell volume $V$ to completely specify
the structure of $\alpha$-Sn, the $\beta$- and $\gamma$-Sn phases
require that we also know the value of $c/a$, the ratio the lattice
constant in the $z$ direction compared to the lattice constant in the
$x-y$ plane.  This means that a static lattice calculation will
determine an energy $U(V,c/a)$.  Fixing $V$ and finding the minimum
energy as a function of $c/a$ will determine the energy $U(V)$ at
that volume.  As the value of $c/a$ which minimizes
(\ref{equ:freeeng}) can change with temperature, properly we should
compute it as a function of of $V, c/a,$ and $T$, and determined the
free energy by minimizing (\ref{equ:freeeng}) in both $V$ and $c/a$ at
fixed $T$.  In practice the change in $c/a$ with volume is so small
during thermal expansion that this is not necessary, so we used the
value of $c/a$ determined by the static VASP calculation at all
temperatures.

Visualizing and analyzing this data was accomplished using third-party
software.  In particular,
\begin{itemize}
\item The crystal structures shown in the next section were plotted
  using Jmol.\cite{hanson:jmol}
\item Phonon calculations in APL were checked comparing the phonon
  frequencies at selected wavelengths along high symmetry lines in the
  Brillouin zone with the frequencies computed by the frozen-phonon
  code FROZSL\cite{Stokes:FROZSL} at the same point.  This helped us
  to determine the appropriate supercell k-point mesh for the simple
  hexagonal $\gamma$-Sn calculations.
\item Some of the experimental phonon frequencies appearing in the
  figures were taken from published graphs.  We used the Engauge
  Digitizer\cite{Mitchell_Engauge_2019} to convert this data into a
  form we could use.  Any errors in the process are ours.
\end{itemize}

\begin{table*}
  \caption{\label{tab:structures} The elemental crystal structures
    investigated in this paper.  As Lonsdaleite does not have a {\em
      Strukturbericht} designation we abbreviate it as ``Lons.''}

  \begin{tabular}{ccccc}
    Common Name & {\em Strukturbericht} & Atoms/Cell & Space Group &
    AFLOW prototype \\
    \hline
    fcc (face-centered cubic) & $A1$ & 1 & $Fm\overline{3}m ~\#225$ &
    \href{\csdb/A\_cF4\_225\_a.html}{A\_cF4\_225\_a}
    \\
    bcc (body-centered cubic) & $A2$ & 1 & $Im\overline{3}m ~\#229$ &
    \href{\csdb/A\_cI2\_229\_a.html}{A\_cI2\_229\_a}
    \\
    hcp (hexagonal close-packed) & $A3$ & 2 & $P6_{3}/mmc ~\#194$ &
    \href{\csdb/A\_hP2\_194\_c.html}{A\_hP2\_194\_c}
    \\
    diamond ($\alpha$-Sn) & $A4$ & 2 & $Fd\overline{3}m ~\#227$ &
    \href{\csdb/A\_cF8\_227\_a.html}{A\_cF8\_227\_a} \\
    $\beta$-Sn & $A5$ & 2 & $I4_{1}/amd ~\#141$ &
    \href{\csdb/A\_tI4\_141\_a.html}{A\_tI4\_141\_a}
    \\
    In (body-centered tetragonal) & $A6$ & 1 & $I4/mmm ~\#139$ &
    \href{\csdb/A\_tI2\_139\_a.In.html}{A\_tI2\_139\_a.In}
    \\
    $\alpha$-Pr (body-centered tetragonal)& $A_{a}$ & 1 & $I4/mmm ~\#139$ &
    \href{\csdb/A\_tI2\_139\_a.alpha-Pa.html}{A\_tI2\_139\_a.alpha-Pa}
    \\
    $\gamma$-Sn (simple hexagonal) & $A_{f}$ & 1 & $P6/mmm ~\#191$ &
    \href{\csdb/A\_hP1\_191\_a.html}{A\_hP1\_191\_a}
    \\
    sc (simple cubic) & $A_{h}$ & 1 & $Pm\overline{3}m ~\#221$  &
    \href{\csdb/A\_cP1\_221\_a.html}{A\_cP1\_221\_a} \\
    Lonsdaleite (hexagonal diamond) & Lons & 4 & $P6_{3}/mmc ~\#194$ &
    \href{\csdb/A\_hP4\_194\_f.html}{A\_hP4\_194\_f}
  \end{tabular}
\end{table*}

\section{\label{sec:structures} Crystal Structures}

We determined the static lattice energy/volume behavior $U(V)$ for tin
using many of the crystal structures observed in the group-IV elements
as well as some close-packed and nearly close-packed elemental
structures.  As there are many notations for these structures it is
useful to describe them here:
\begin{itemize}
\item The common names of the structures ({\em e.g.}, fcc,
  bcc, diamond or $\alpha$-Sn, $\beta$-Sn, simple hexagonal or
  $\gamma$-Sn, {\em etc.}).
\item Elemental structures are most easily described by their {\em
  Strukturbericht} designations, as given in the original works
  published by Ewald {\em et al.},\cite{Ewald_et_al_Strukturbericht}
  or the extensions proposed by
  Smithells.\cite{Smithells_Metals_II_1955} The exception here is
  Lonsdaleite,\cite{Yoshiasa_Japanese_J_App_Phys_42_1694_2003}
  hexagonal diamond, which has no \strb{} entry.  We will primarily
  use the \strb{} labels in graphs to avoid clutter.
\item For high-throughput calculations it is useful to have a label
  which allows the user to compactly specify the structure.  Since we
  are using AFLOW, we use the AFLOW prototype
  label,\cite{Mehl_Comp_Mat_Sci_136_S1_2017} which uniquely specifies
  the stoichiometry, space group, and Wyckoff positions of the
  structure.  Thus the diamond ($A4$, $\alpha$-Sn) structure is
  denoted A\_cF8\_227\_a, denoting one type of atom (A), a
  face-centered cubic primitive cell with eight atoms in the
  conventional cell (Pearson symbol cF8), in space group \#227
  ($Fd\overline{3}m$) with the atoms at the (8a) Wyckoff position.
\end{itemize}

More details about the structures, including the above information and
a full description of the primitive lattice vectors and basis vectors
can be found in the Library of Crystallographic
Prototypes,\cite{Mehl_Comp_Mat_Sci_136_S1_2017,Hicks_CMS_161_S1_2019}
available online at \href{\csdb}{\csdb}.  The Library also allows the
user to generate structure files for use as input in a wide variety of
electronic structure codes, including the POSCAR files for these
AFLOW/VASP calculations.

Table~\ref{tab:structures} describes all of the structures used here,
including the common name, {\em Strukturbericht} label, space group,
AFLOW prototype, and a link to the corresponding entry in the Library
of crystallographic prototypes.

The face-centered cubic ($A1$), body-centered cubic ($A2)$, and both
body-centered tetragonal structures ($A6$, $A_{a}$) can all be derived
from one another by stretching or compressing the primitive cell along
the (001) direction, with the $A6$ structure having a $c/a$ ratio
close to the $A1$ structure, and $A_{a}$ near $A2$.

\begin{figure}
  \includegraphics[width=8cm]{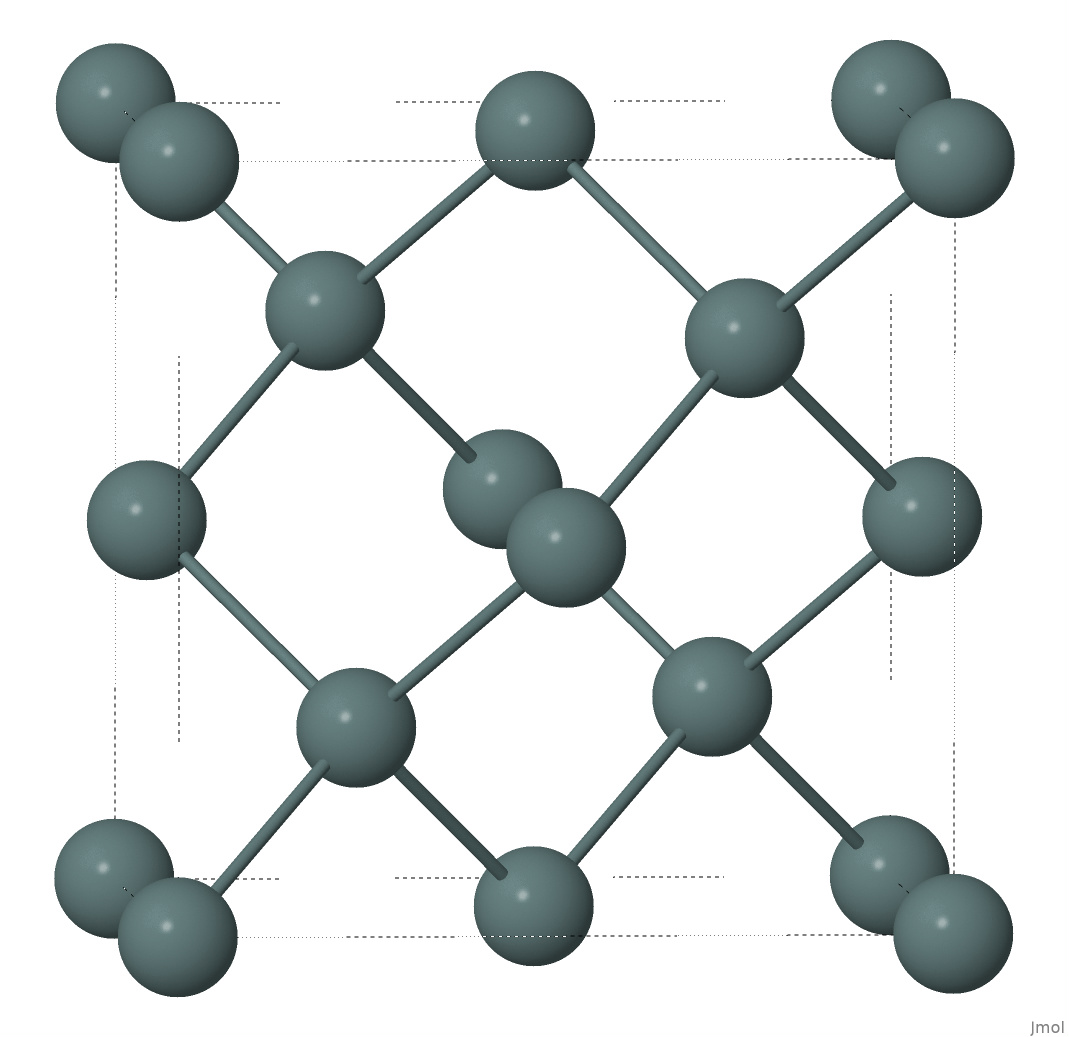}
  \includegraphics[width=8cm]{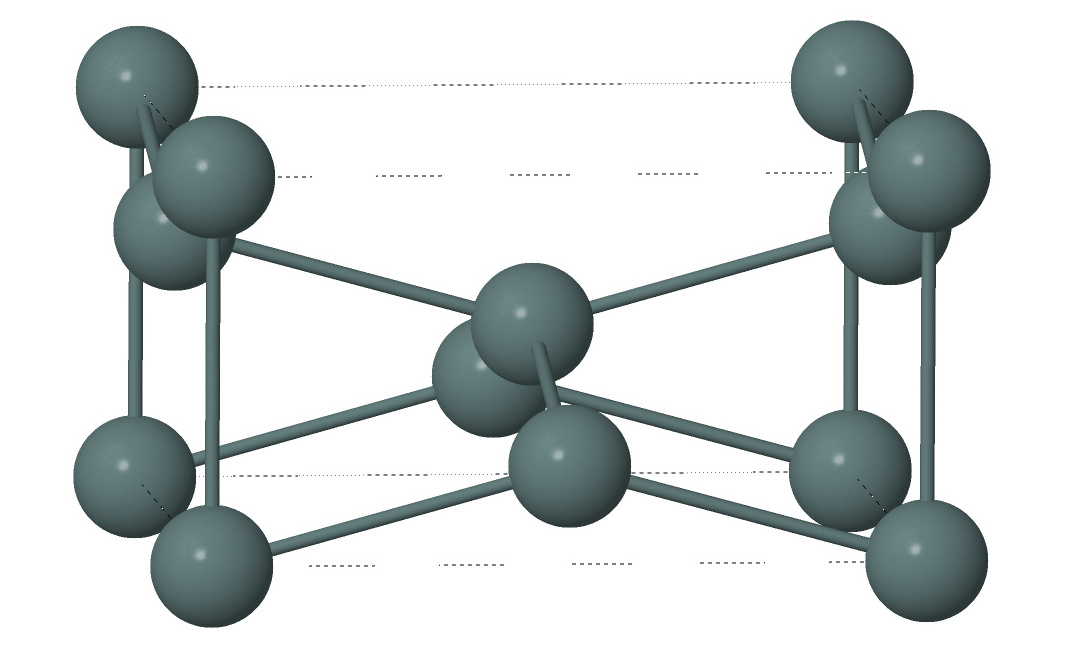}
  \includegraphics[width=8cm]{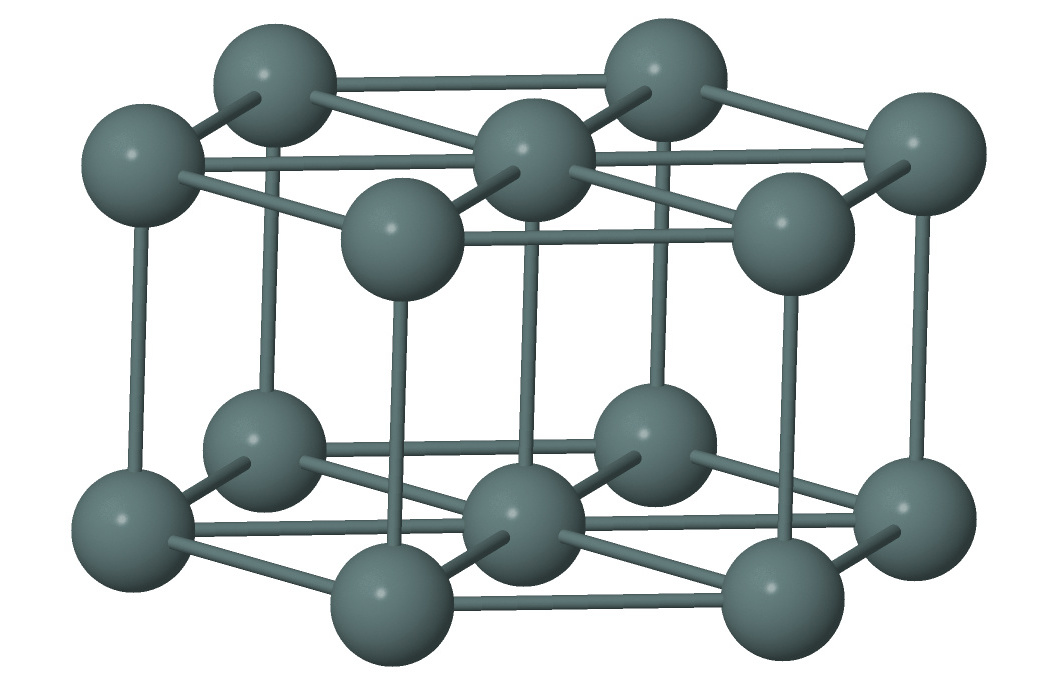}
  \caption{\label{fig:structures} The three low-energy structures of
    tin, drawn approximately to scale. Top: $\alpha$-Sn, \strb{} $A4$
    (gray tin, diamond structure), AFLOW Designation A\_cF8\_227\_a.
    Middle: $\beta$-Sn, \strb{} $A5$ (white tin), AFLOW Designation
    A\_tI4\_141\_a.  Bottom: simple hexagonal $\gamma$-Sn, \strb{}
    $A_{f}$, AFLOW Designation A\_hP1\_191\_a.  The conventional cells
    are shown for the cubic $\alpha$-Sn and tetragonal $\beta$-Sn.
    The $\gamma$-Sn figure contains three primitive cells to show the
    hexagonal structure.}
\end{figure}

The structures of most interest in this work are $\alpha$-Sn (diamond
structure, gray tin, or $A4$), $\beta$-Sn (white tin or $A5$) and
simple hexagonal $\gamma$-Sn ($A_{f}$) structures.  These are shown in
Figure~\ref{fig:structures}.  The $\beta$-Sn structure can be obtained
from $\alpha$-Sn by compressing along the (001) axis of the diamond
crystal.  A possible transition path from $\beta$-Sn to $\gamma$-Sn
was found by Needs and Martin.\cite{Needs_PRB_30_5390_1984}

\subsection{\label{subsec:Af} The $\gamma$-Sn structure}

While the $\alpha$- and $\beta$-Sn structures are well known, the
$\gamma$-Sn structure is not.
Smithells,\cite{Smithells_Metals_II_1955} apparently referencing
Raynor and Lee,\cite{Raynor_Acta_Met_2_616_1954} designated
HgSn$_{10}$ as the prototype for \strb{} designation $A_{f}$, with
hexagonal space group $P6/mmm ~\#191$ and one atom per unit cell
located at the (1a) Wyckoff position.  This can only be achieved if
the mercury and tin atoms are randomly placed on the (1a) site.
Alloys of tin with 5-20\% of cadmium, indium, lead and mercury also
exhibit this structure, which is generally referred to as the
$\gamma$-phase.\cite{Kane_Acta_Met_14_605_1966,Ivanov_J_Phys_F_17_1925_1987,Ivanov_Physica_B_174_79_1991}
Parth\'{e} {\em et al.}\cite{Parthe_Gmelin_Handbook_1993} also list
In$_{0.45}$Bi$_{0.55}$ under the $A_{f}$ designation.

Though it is not the ground state of any element, the simple hexagonal
phase ($\gamma$-Sn) is observed at high pressures in
silicon\cite{Needs_PRB_30_5390_1984} and
germanium.\cite{Ackland_RPP_64_483_2001} When studying this phase it
was found\cite{Needs_PRB_30_5390_1984} that the simplest possible
transition between $\beta$-Sn and $\gamma$-Sn is described by
distorting either of the phases into a body-centered orthorhombic
crystal, space group $Imma ~ \#74$, with the atoms at the (4e) Wyckoff
position.  This ``Needs-Martin'' path can be described using a
body-centered orthorhombic unit cell, space group $Imma \#74$, with
atoms occupying the (4e) Wyckoff positions, locating the atoms at
\begin{equation}
  \label{equ:needsmartin}
  \vec{b}_{\pm} = \pm (1/4 b \hat{y} + z c \hat{z}) ~ .
\end{equation}
When $a = b$ and $z = 1/8$ this becomes the $\beta$-Sn structure.  The
$\gamma$-Sn structure is reached when $z = 1/4$, $a = 2 c_{hex}$, $b =
\sqrt{3} a_{hex}$, and $c = a_{hex}$, where $a_{hex}$ and $c_{hex}$
are the lattice constants of the hexagonal structure.  Note that in
this case the primitive cell contains two of the hexagonal primitive
cells.  This gives us a simple path to study the $\alpha\mbox{-Sn}
\leftrightarrow \beta\mbox{-Sn} \leftrightarrow \gamma\mbox{-Sn}$
transitions.

An elemental $\gamma$-Sn phase, occurring above 435K, was apparently
described in texts around 1960, ``but it is no longer
referenced.''\cite{Kubiak_JLCM_116_307_1986} In 1985
Kubiak\cite{Kubiak_JLCM_116_307_1986} found that a structure he called
the $\gamma$ phase appeared after heating single crystal $\beta$-Sn in
air at 450K for one week.  He described this structure as having space
group $Cmmm ~\#65$, with two atoms in the conventional orthorhombic
cell located on the (2a) Wyckoff position, and lattice parameters
$(a,b,c) = $(5.8308\,\AA, 3.181\,\AA, 2.9154\,\AA).  This structure is
extremely close to simple hexagonal $A_{f}$, and when we run
electronic structure calculations starting with Kubiak's $\gamma$-Sn
structure it always relaxes to the simple hexagonal $A_{f}$ structure.
Given this we will only consider the hexagonal structure in our
calculations below, and refer to it as both $\gamma$-Sn and $A_{f}$.

The $\gamma$-Sn phase has been studied by Wehinger {\em et
  al.},\cite{Wehinger_JPCM_26_115401_2014} who found that it was
energetically similar to $\beta$-Sn using the LDA.  They did not
address its thermal behavior, nor did they discuss the relationship of
the two high-temperature phases with the ground state $\alpha$-Sn
structure.

The orthorhombic elemental $\gamma$-Sn phase also can be stabilized in
tin nanoparticles and nanowires.\cite{Hormann_APL_107_123101_2015} We
will not address this work here.

\section{\label{sec:dftresults} Density functionals used to study tin}

In our study of the the tungsten-nitrogen
system\cite{Mehl_PRB_91_184100_2015} we found that the predicted
ground state structure of a compound can change with the choice of
density functional.  Given the small energy difference between tin
phases it is quite possible that different functionals will give
different ground state structures.  In this section we look at
possible structures of tin with a variety of density functionals, all
available in VASP:

\begin{itemize}
\item The Local Density Approximation
  (LDA),\cite{Hedin_J_Phys_C_4_2064_1971,Ceperley_PRL_45_566_1980,Perdew_PRB_23_5048_1981}
  which derives the Kohn-Sham potential\cite{Kohn_PR_140_A1133_1965}
  at a given point in space using only the charge density at that point.
  This is well-known to underestimate equilibrium lattice volumes.
\item Generalized-Gradient Approximation (GGA) functionals, where the
  Kohn-Sham potential depends on the local charge density and its
  local gradient.  These include
  \begin{itemize}
  \item Perdew-Burke-Ernzerhof (PBE),\cite{Perdew_PRL_77_3865_1996}
    perhaps the most widely used GGA.  This generally overestimates
    equilibrium lattice volumes.
  \item Perdew-Burke-Ernzerhof revised for solids
    (PBEsol),\cite{Perdew_PRL_100_136406_2008} a modification of PBE
    optimized for solids rather than atoms.
  \item Armiento-Mattsson (AM05),\cite{Armiento_PRB_72_085108_2005}
    designed to describe surfaces, but which has proved to be very
    accurate for solids.\cite{Mattsson_JCP_128_084714_2008}
  \end{itemize}
\item Meta-GGA functionals, which depend on the orbital kinetic energy
  density as well as the charge density and its gradient.  VASP
  provides
  \begin{itemize}
  \item Tao-Perdew-Staroverov-Scusera
    (TPSS),\cite{Tao_PRL_91_146401_2003} designed to be correct for
    one- and two-electron systems and systems with slowly varying charge
    densities.
  \item  ``Revised'' TPSS (revTPSS),\cite{Perdew_PRL_103_206403_2009}
    which includes the second-order gradient expansion for exchange.
  \item ``Made-simple'' functionals (MS0, MS1,
    MS2),\cite{Sun_JCP_137_051101_2012,Sun_JCP_138_044113_2013} which
    have empirical parameters.
  \item M06-L,\cite{Zhao_JCP_125_194101_2006} optimized for main-group
    and transition metal chemistry.
  \item ``Strongly Constrained and Appropriately Normed''
    (SCAN),\cite{Sun_PRL_115_036402_2015} which satisfies all known
    constraints on the exact density functional with no adjustable
    parameters.  Since the SCAN functional properly describes both
    covalent and metallic bonding in
    silicon,\cite{Sun_Nature_Chem_8_831_2016} it should be able to
    describe similar behavior in tin.
  \end{itemize}
\end{itemize}

These functionals have been tested over a variety of
datasets,\cite{Mattsson_JCP_128_084714_2008,Csonka_PRB_79_155107_2009,Sun_PRB_84_035117_2011}
but to our knowledge no test has been made on the accuracy of any of
these functionals in describing the tin phase transition.

The remainder of this section discusses the behavior of tin as
predicted by high-throughput calculations using the functionals
described above.  We first look at the energy-volume curves for tin in
the phases described in Section \ref{sec:structures} and
Table~\ref{tab:structures}.  If a functional appears to give a
reasonable description of the behavior of tin we will look at its
predictions of tin's thermal expansion and the $\alpha$-$\beta$
transition in Section~\ref{sec:thermal}.

\begin{figure}
  \includegraphics[width=0.5\textwidth]{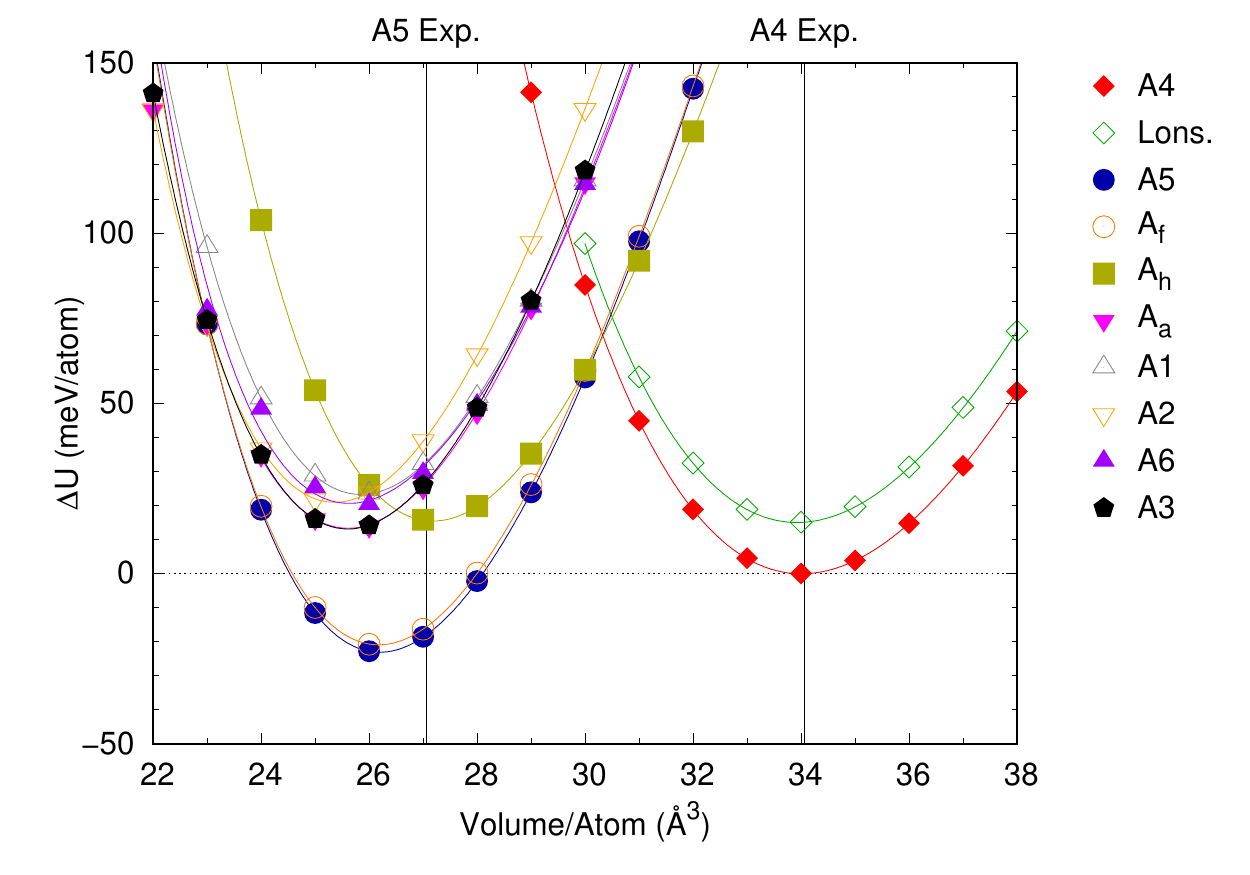}
  \caption{\label{fig:ldaev} Static lattice energy-volume curves for
    the tin structures discussed in Section~\ref{sec:structures} as
    predicted by AFLOW/VASP using the LDA functional.  Structural
    notation is from Table~\ref{tab:structures}.  We plot $\Delta$U,
    the change in energy per atom compared to the equilibrium energy
    of the $\alpha$-Sn ($A4$) structure.  The lines labeled ``A5
    Exp.'' and ``A4 Exp.''  represent the experimental volume of
    $\beta$-Sn ($A5$) at 298K\cite{Deshpande_Acta_Cryst_14_355_1961}
    and $\alpha$-Sn ($A4$) at 90K\cite{Price_PRB_3_1268_1971}
    respectively.}
\end{figure}

\subsection{\label{subsec:lda} The Local Density Approximation (LDA)}

The energy-volume curves predicted by the Local Density Approximation
(LDA) are shown in Figure~\ref{fig:ldaev}.

As is usual with the LDA the predicted equilibrium volume for
$\beta$-Sn is about 4\% below the experimental volume.  Somewhat
surprisingly the equilibrium volume for $\alpha$-Sn is approximately
equal to the low-temperature experimental volume.

As one expects there is a large equilibrium volume difference between
$\alpha$- and $\beta$-Sn, in agreement with experiment.  Lonsdaleite
(``Lons.''  in the figure), the hexagonal diamond structure, is
correctly above $\alpha$-Sn ($A4$), and $\gamma$-Sn ($A_{f}$) is above
$\beta$-Sn ($A5$).  We see that the LDA overbinds both $\beta$-Sn and
$\gamma$-Sn with respect to $\alpha$-Sn, contrary to experiment,
predicting a $\beta$-Sn ground state.  In addition, $\gamma$-Sn is
nearly degenerate with $\beta$-Sn, which, while not completely ruled
out by experiment seems somewhat unlikely.

The prediction of $\beta$-Sn as the ground state within the LDA was
also observed by Christensen and
Methfessel\cite{Christensen_PRB_48_5797_1993} using the LMTO-ASA
method.  They found an equilibrium energy difference of 5\,meV/atom,
substantially less than our value of 20\,meV/atom, but in neither case
will we get the experimentally observed thermal transition.

\begin{figure}
  \includegraphics[width=0.5\textwidth]{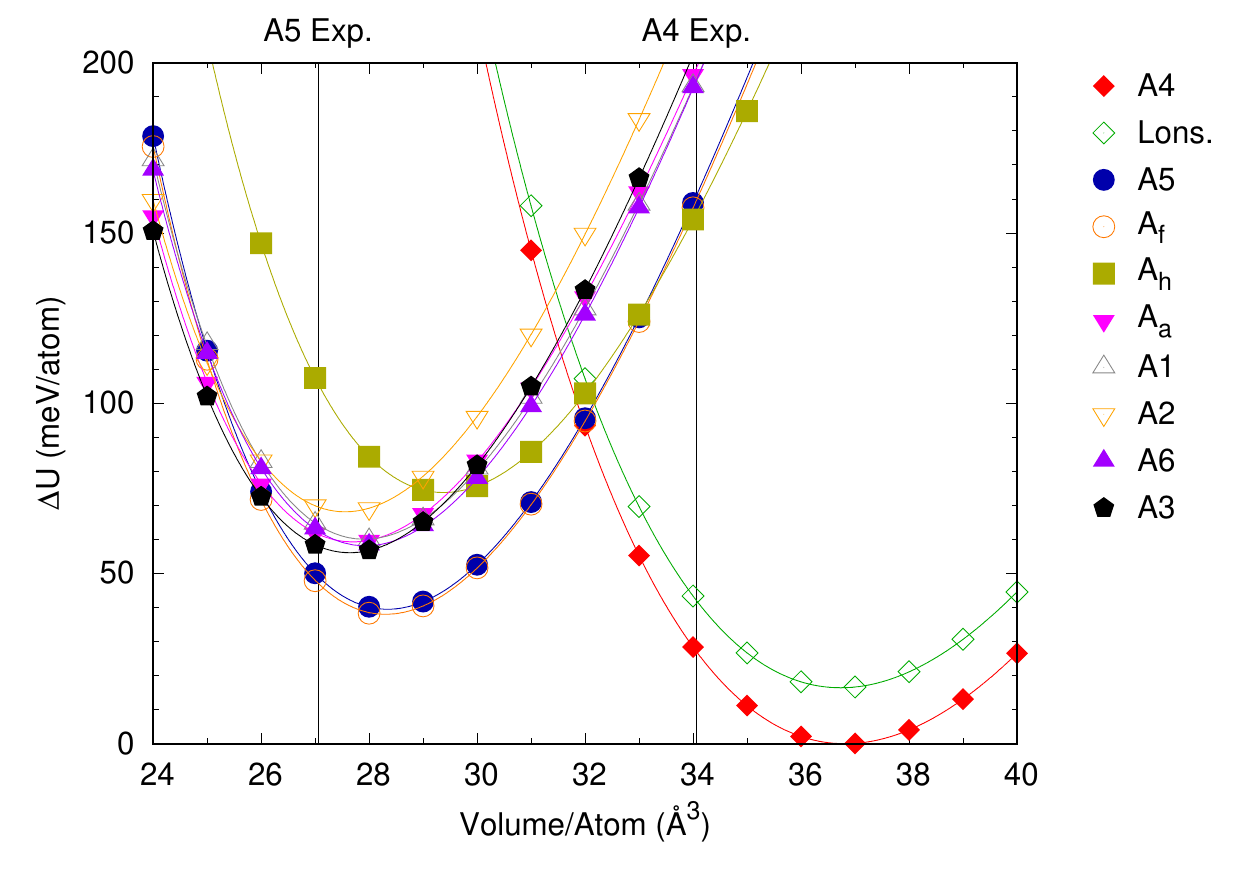}
  \caption{\label{fig:pbeev} Static lattice energy-volume curves for
    for the tin structures discussed in Section~\ref{sec:structures}
    as predicted by AFLOW/VASP using the PBE functional.  The notation
    is identical to that in Figure~\ref{fig:ldaev}.}
\end{figure}

\subsection{\label{subsec:pbe} The Perdew-Burke-Ernzerhof Generalized
  Gradient Functional (PBE)}

The results for the PBE functional are shown in
Figure~\ref{fig:pbeev}.  As is usual with the PBE, the predicted
equilibrium volumes are 3-10\% above the experimental values.

The PBE is an improvement over LDA in that it predicts that the
equilibrium energy of the $\beta$-Sn phase is 40\,meV above the
$\alpha$-Sn, in agreement with previous
work.\cite{Legrain_JCP_143_204701_2015} As with LDA, the $\beta$- and
$\gamma$-Sn phases are nearly degenerate, but in the PBE the
$\gamma$-phase is actually lower in energy than the $\beta$-Sn.  Since
$\gamma$-Sn is only observed at 450K and above (if at all), this is
either an error of the PBE or the $\beta$-Sn phase must be stabilized
by vibrational effects.

\subsection{\label{subsec:pbesol} The Perdew-Burke-Ernzerhof Generalized
  Gradient Functional Revised for Solids (PBEsol)}

\begin{figure}
  \includegraphics[width=0.5\textwidth]{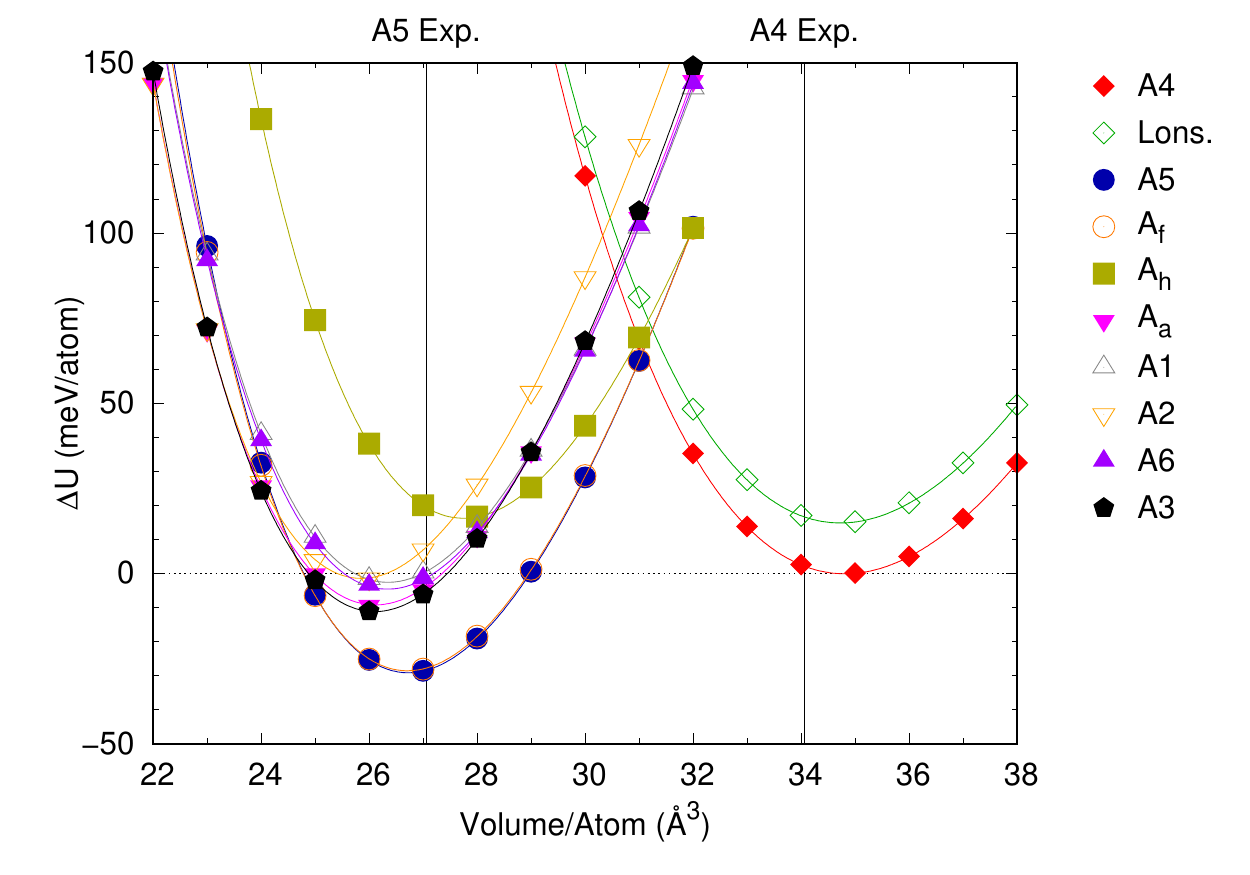}
  \caption{\label{fig:pbesolev} Static lattice energy-volume curves
    for for the tin structures discussed in
    Section~\ref{sec:structures} as predicted by AFLOW/VASP using the
    PBEsol functional.  The notation is identical to that in
    Figure~\ref{fig:ldaev}.}
\end{figure}

The results for the PBEsol functional are shown in
Figure~\ref{fig:pbesolev}.  Since this functional was designed to give
better equilibrium volumes than PBE, it is not surprising that the
equilibrium volumes for the $\alpha$-, $\beta$-, and $\gamma$- phases
are between those predicted by the LDA and PBE functionals.

In other respects the PBEsol results are slightly worse than those
found by the LDA.  The $\beta$-Sn phase is even more bound compared to
the $\alpha$- phase, and the $\beta$- and $\gamma$- phases are closer
together and nearly indistinguishable on the plot.

\subsection{\label{subsec:am05} The Armiento-Mattsson Generalized
  Gradient Functional (AM05)}

\begin{figure}
  \includegraphics[width=0.5\textwidth]{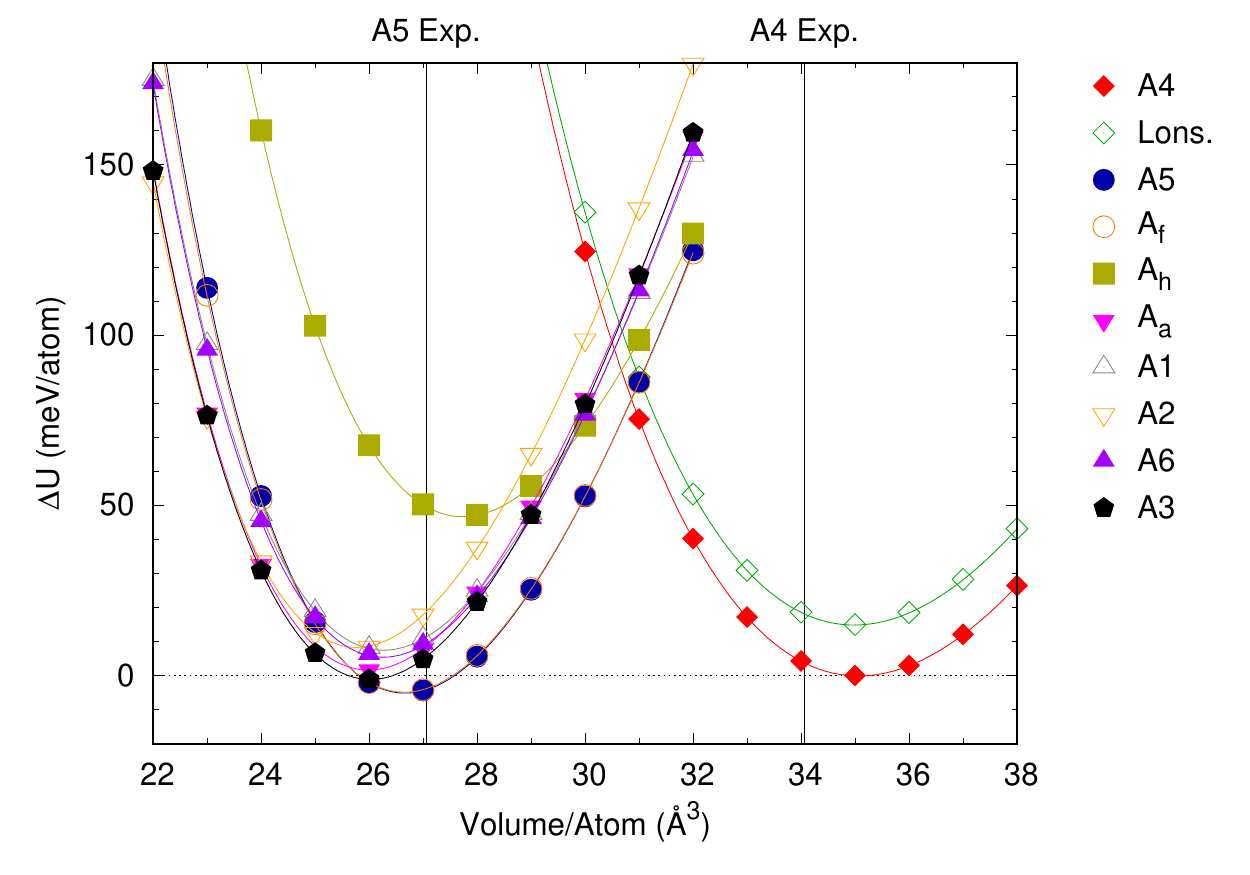}
  \caption{\label{fig:am05ev} Static lattice energy-volume curves for
    for the tin structures discussed in Section~\ref{sec:structures}
    as predicted by AFLOW/VASP using the AM05 functional.  The
    notation is identical to that in Figure~\ref{fig:ldaev}.}
\end{figure}

The results for the AM05 functional are shown in
Figure~\ref{fig:am05ev}.  The equilibrium properties are similar to
PBEsol, except that the $\beta$- and $\gamma$-Sn phases are less
weakly bound compared to diamond than they are in LDA or PBEsol
calculations.

It is interesting to look at the predictions for the close-packed
($A1$, $A3$) and near close-packed ($A2$, $A6$, $A_{a}$) phases in
this study.  The previous functionals predicted that these phases had
minimum energies 20-40\,meV/atom above $\beta$- and $\gamma$-Sn.  Here
the energy difference is only about 5\,meV/atom.  This small energy
difference leads to a prediction of a transition from $\beta$-Sn to
body-centered cubic Sn at 1\,GPa, far below the observed experimental
transition at 35\,GPa.\cite{Olijnyk_J_Phys_Coll_45_C8_153_1984}

\subsection{\label{subsec:metagga} Meta-GGA Functionals (except SCAN)}

\begin{figure}
  \includegraphics[width=0.5\textwidth]{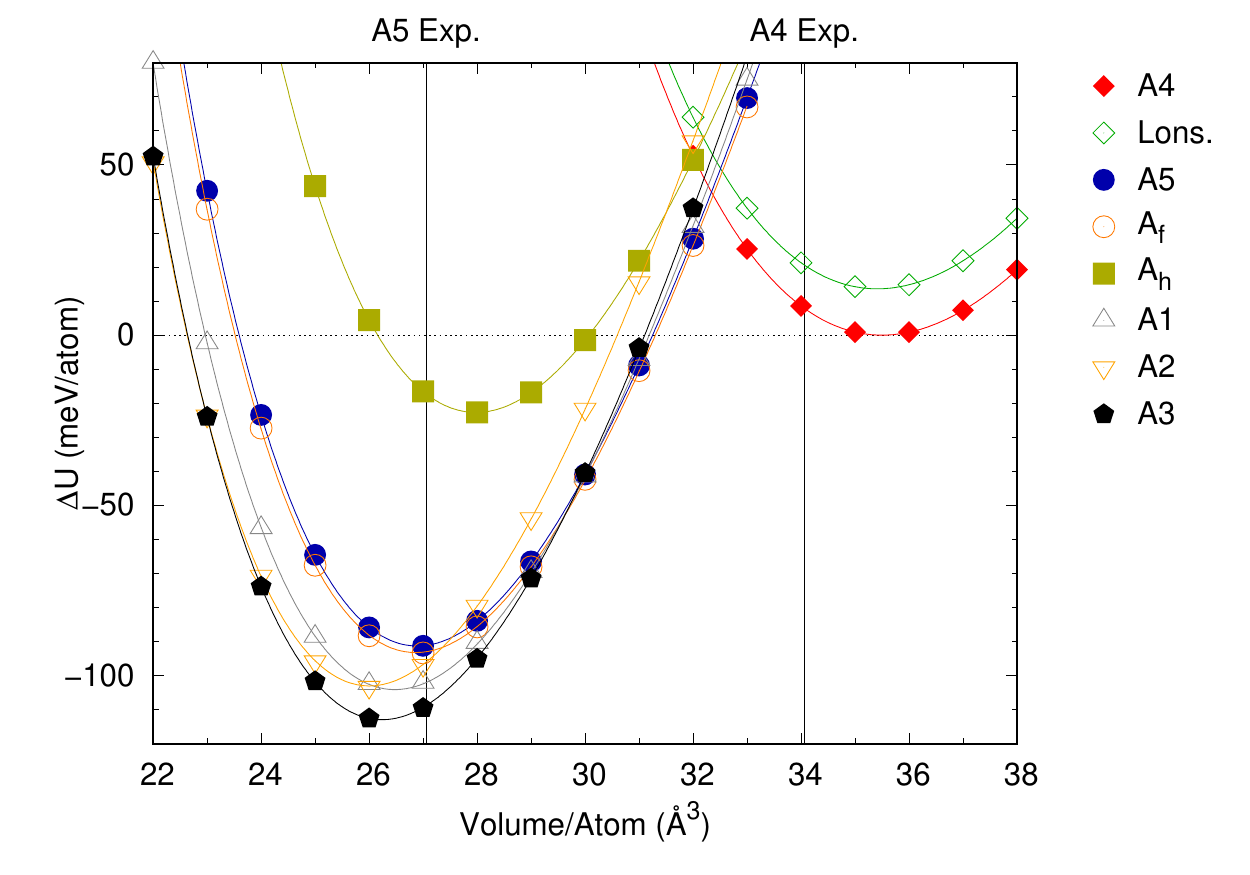}
  \caption{\label{fig:RTPSS} Static lattice energy-volume curves for
    for most of the tin structures discussed in
    Section~\ref{sec:structures} as predicted by AFLOW/VASP using the
    revTPSS functional.  The notation is identical to that in
    Figure~\ref{fig:ldaev}.  We did not do all of the structures in
    for this or the other non-SCAN meta-GGAs because of the time
    involved in the calculations.  All of the non-SCAN meta-GGAs
    listed in Section \ref{sec:dftresults} significantly overbind the
    close-packed phases with respect to $\alpha$-, $\beta$-, and
    $\gamma$-Sn.}
\end{figure}

All of the meta-GGA functionals (TPSS, revTPSS, MS0/1/2, M06-L), with
the exception of the SCAN functional discussed below, significantly
overbind the close-packed fcc ($A1$) and hcp ($A3$) structures, as
well as the nearly close-packed bcc ($A2$) structure compared to
$\alpha$-, $\beta$-, and $\gamma$-Sn.  Calculations using these
functionals are significantly more time-consuming than LDA or GGA
calculations, so once we realized this, we screened the functionals by
looking at the energy difference between the hcp ($A3$) and
$\alpha$-Sn ($A4$) phases.  We did look at most of our target
structures with the revTPSS functional, and show these results in
Figure~\ref{fig:RTPSS}.

These functionals have been optimized for non-covalent
interactions\cite{Zhao_JCP_125_194101_2006} so it is not surprising
that they do not describe the energetics of the covalently-bonded
$\alpha$-Sn phase particularly well.

\subsection{\label{subsec:scan} The Strongly Constrained and
  Appropriately Normed (SCAN) meta-GGA}

\begin{figure}
  \includegraphics[width=0.5\textwidth]{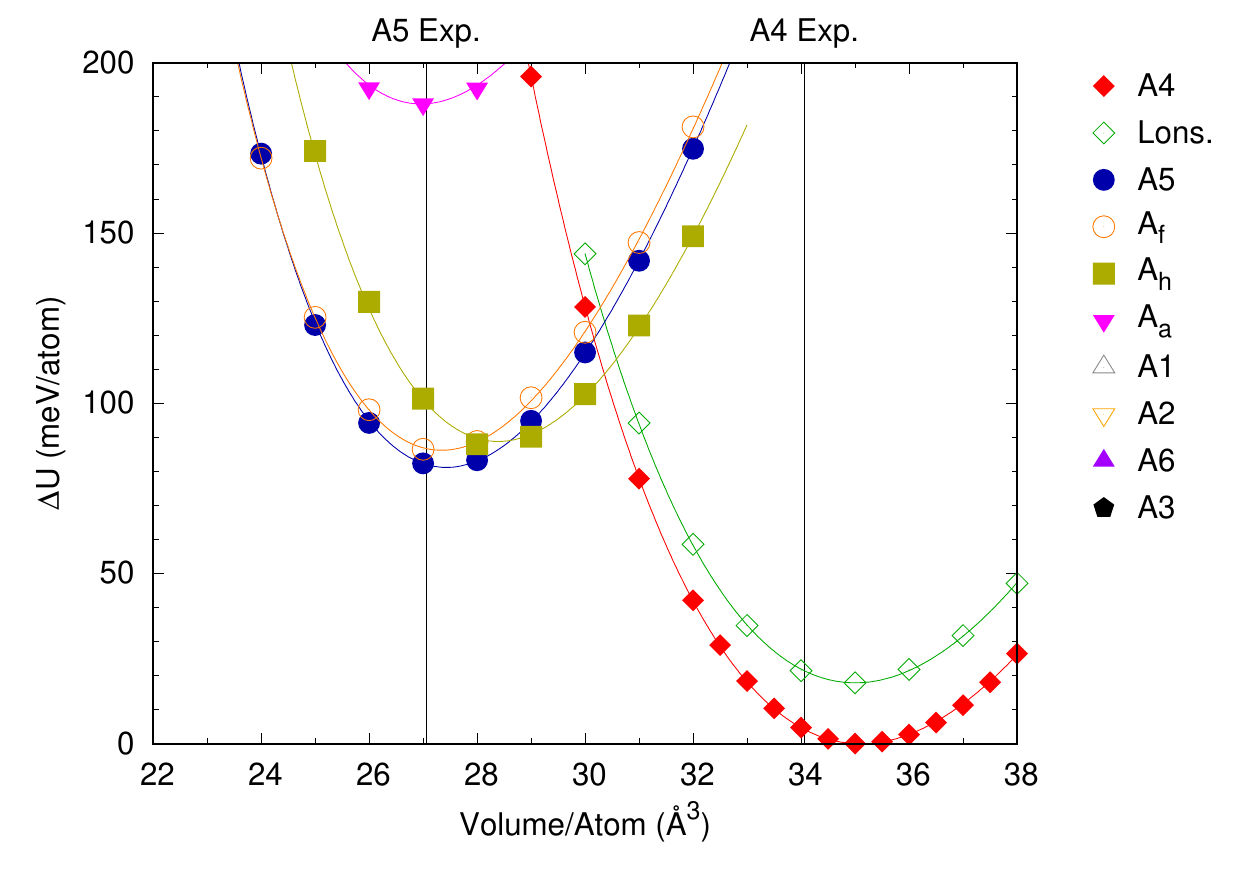}
  \caption{\label{fig:scanev} Static lattice energy-volume curves for
    for the tin structures discussed in Section~\ref{sec:structures}
    as predicted by AFLOW/VASP using the SCAN functional.  The
    notation is identical to that in Figure~\ref{fig:ldaev}.  Structures
    not shown ($A1, A2, A3, A6$) are above the $\Delta U =
    200$\,meV/atom limit of the graph.}
\end{figure}

Unlike meta-GGAs such as M06-L, the SCAN functional is non-empirical
and is designed ``to satisfy all 17 exact constraints appropriate to a
semilocal functional.''\cite{Sun_Nature_Chem_8_831_2016} As shown in
Figure~\ref{fig:scanev}, SCAN predicts the correct ordering of the
major tin phases, $U(\alpha\mbox{-Sn}) < U(\beta\mbox{-Sn}) <
U(\gamma\mbox{-Sn})$.  The simple cubic ($A_{h}$) phase is very low
compared to other calculations, while the close-packed and nearly
close-packed phases barely make the graph, with only the body-centered
tetragonal $\alpha$-Pr ($A_{a}$) phase within 200\,meV of $\alpha$-Sn.

\begin{table*}
  \caption{\label{tab:equprop}Predicted equilibrium properties of
    $\alpha$-, $\beta$-, and $\gamma$-Sn for the density functionals
    described in Section~\ref{sec:dftresults}, as well as the LDAU
    calculations described in Section~\ref{sec:hubbard} These
    calculations were made by allowing VASP to fully relax each unit
    cell, so the equilibrium volumes and energies will differ slightly
    from the Birch fit derived quantities used elsewhere in this
    paper.  $\Delta$U$_{\alpha\beta}$ and $\Delta$U$_{\alpha\gamma}$
    represent the equilibrium energy difference between $\beta$-Sn or
    $\gamma$-Sn and $\alpha$-Sn, respectively.  A positive number
    indicates that $\alpha$-Sn is lower in energy.  We also include
    experimentally measured lattice constants for comparison.  The
    experimental lattice constants for $\gamma$-Sn are for the alloy
    with stoichiometry
    Sn$_{0.8}$In$_{0.2}$.\cite{Ivanov_J_Phys_F_17_1925_1987}}
  \begin{tabular}{cccccccc}
    & $\alpha$-Sn & \multicolumn{2}{c}{$\beta$-Sn} &
    \multicolumn{2}{c}{$\gamma$-Sn} & $\Delta$U$_{\alpha\beta}$
    & $\Delta$U$_{\alpha\gamma}$ \\
    & a (\AA) & a (\AA) & c (\AA) & a (\AA) & c (\AA) & (eV/atom) &
    (eV/atom) \\
    \hline
    Exp. (90K) & 6.483\cite{Price_PRB_3_1268_1971} \\
    Exp. (100K) & & 5.815\cite{Rowe_PRL_14_554_1965} & 3.164\cite{Rowe_PRL_14_554_1965} \\
    Exp. (296K) & 6.491\cite{Ihm_PRB_23_1576_1981} &
    5.832\cite{Rowe_PRL_14_554_1965} & 3.183\cite{Rowe_PRL_14_554_1965}
    & 3.216\cite{Ivanov_J_Phys_F_17_1925_1987} &
    2.998\cite{Ivanov_J_Phys_F_17_1925_1987} \\
    Exp. (300K) & & 5.3815\cite{Price_Proc_Roy_Soc_Lon_A_300_25_1967} &
    3.1828\cite{Price_Proc_Roy_Soc_Lon_A_300_25_1967} \\
    \hline
    LDA & 6.4787 & 5.7823 & 3.1325 & 3.1822 & 2.9844 & -0.023 & -0.021
    \\
    PBE & 6.6528 & 5.9334 & 3.2187 & 3.2645 & 3.0678 & 0.040 & 0.039
    \\
    PBEsol & 6.5296 & 5.8193 & 3.1550 & 3.2024 & 3.0052 &
    -0.029 & -0.029 \\
    AM05 & 6.5453 & 5.8135 & 3.1554 & 3.2000 & 3.0030 & -0.005 &
    -0.005 \\
    revTPSS & 6.5501 & 5.8124 & 3.1502 & 3.1962 & 3.0034 & -0.092 &
    -0.094 \\
    SCAN & 6.4235 & 5.7286 & 3.1116 & 3.1572 & 2.9551 & 0.081 & 0.086
    \\
    LDA (LDAU, U$_{p}$ = -0.80) & 6.4235 & 5.7286 & 3.1116 & 3.1572 & 2.9551
    & 0.020 & 0.028 \\
    PBE (LDAU, U$_{p}$ = 0.20) & 6.6751 & 5.9541 & 3.2271 & 3.2733 &
    3.0805 & 0.025 & 0.022
  \end{tabular}

\end{table*}

Table~\ref{tab:equprop} provides a brief summary of the equilibrium
properties of the three tin phase for each choice of DFT, along with
the static energy difference between the three phases.  The SCAN
functional is the only one which correctly predicts the energy
relationship $U(\alpha\mbox{-Sn}) < U(\beta\mbox{-Sn}) <
U(\gamma\mbox{-Sn})$.  In that sense it is better than any of the
other functionals studied.  Unfortunately the relative energy
$U(\beta\mbox{-Sn}) - U(\alpha\mbox{-Sn})$ is approximately
80\,meV/atom, significantly larger than the 10-40\,meV/atom suggested
by the 286K transition
temperature,\cite{Pavone_PRB_57_10421_1998,Houben_PRB_100_075408_2019}
and like most of the functionals discussed here it overestimates the
equilibrium volumes for both gray and white tin.

\section{\label{sec:thermal} Phonons and Thermodynamics}

Of all the density functionals studied in
Section~\ref{sec:dftresults}, only two predict the $\beta$-Sn phase to
have a higher energy than the $\alpha$-Sn phase: PBE and SCAN.  The
PBE functional has the apparent deficiency that the $\gamma$-Sn phase
is lower in energy than the $\beta$-phase, but the energy difference
is small and perhaps zero-point and/or thermal effects will stabilize
$\beta$-Sn at finite temperatures.  The AFLOW APL module allows us to
determine the zero-point and temperature-dependent free energy for all
three of the tin phases.  We will begin with the PBE functional.

The procedure is as follows:
\begin{itemize}
\item Determine the total energy of the three possible phases of tin
  as a function of volume using AFLOW/VASP.
\item Use the harmonic phonon module (APL) of AFLOW to determine the
  phonon spectra for each structure and volume.
\item The APL module automatically determines the vibrational free
  energy (\ref{equ:freeeng}) as a function of temperature for each of
  these structures and volumes, including the zero-point energy.
  Since the variation of $c/a$ for the $\beta$- and $\gamma$-phases is
  small, we will ignore change in $c/a$ with temperature and use the
  $c/a$ found to minimize the static energy at each volume for all
  temperatures at the volume.
\item For a given temperature, determine the volume which minimizes
  the free energy using the Birch fit (\ref{equ:birch}), generating
  the free energy $F(T)$ and equilibrium volume $V(T)$.
\item Determine the averaged linear expansion coefficient of each structure
  using the relationship
  \begin{equation}
    \label{equ:alpha}
    \alpha(T) = \frac1{3 V(T)} \frac{dV}{dT}(T) ~ ,
  \end{equation}
  where $a(T)$ is the lattice constant of the crystal.  For the
  tetragonal $\beta$- and hexagonal $\gamma$-phases $\alpha(T)$ will
  be the average of the linear expansion coefficients in the $x$-,
  $y$-, and $z$-directions.
\item Compare the values of the free energy $F(T)$ for each structure
  to determine the equilibrium structure as a function of temperature.
\end{itemize}

\subsection{\label{subsec:pbephonons} Phonon Spectra}

We must first ask if the APL module correctly describes the
vibrational behavior of tin.  Fortunately experimental phonon spectra
for all three phases are available.

\begin{figure}
  \includegraphics[width=0.5\textwidth]{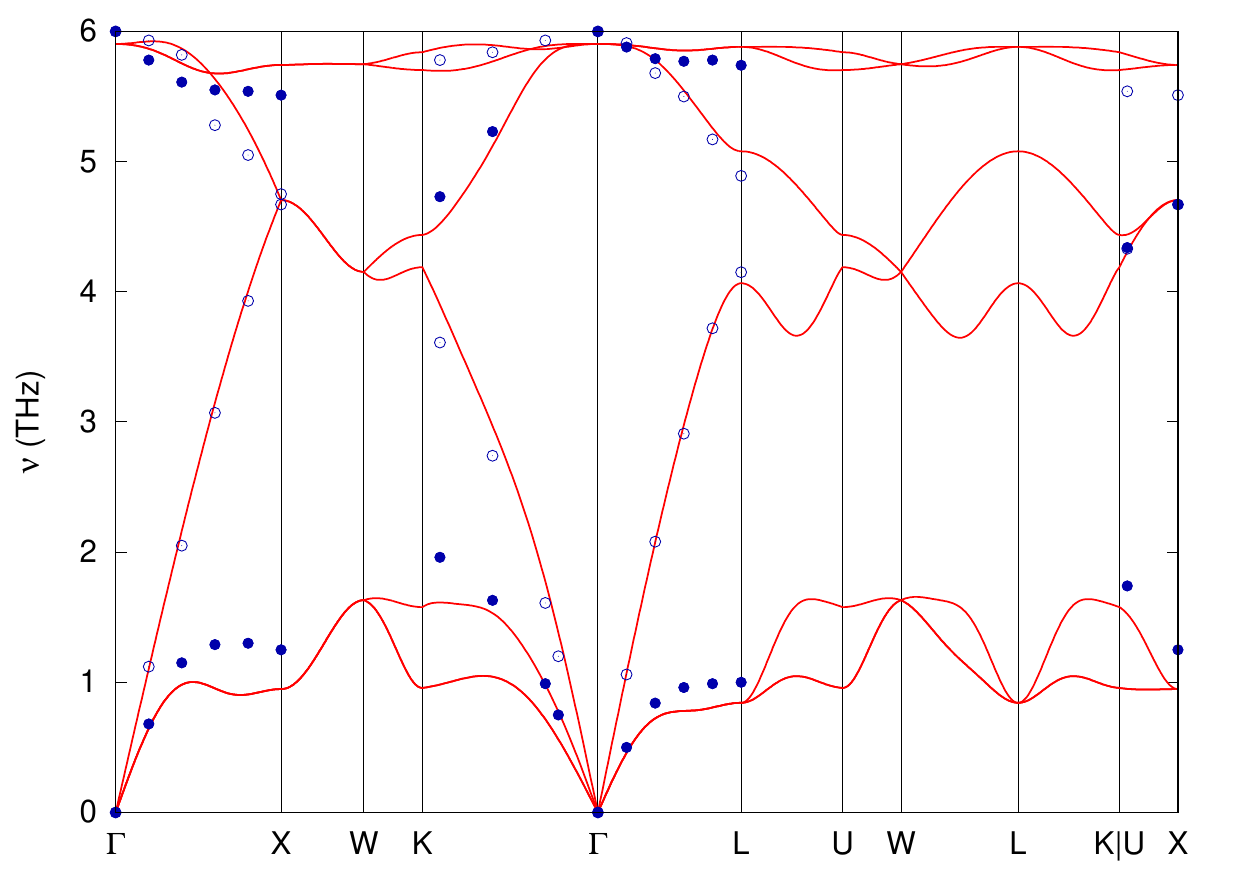}
  \caption{\label{fig:A4PBEph} Transverse (solid circles) and
    longitudinal (open circles) phonon frequencies for $\alpha$-Sn
    measured at 90K by Price {\em et al.},\cite{Price_PRB_3_1268_1971}
    compared to the PBE phonon spectra at the primitive cell volume of
    68\AA$^3$ calculated using the APL module of AFLOW.  The
    high-symmetry paths through the face-centered cubic Brillouin zone
    are defined by Setyawan and
    Curtarolo.\cite{Setyawan_CMS_49_299_2010} Note that Price {\em et
      al.} only determined the frequencies of one of the transverse
    branches along the $\Gamma-K$ and $U-X$ directions (the $\Sigma$
    line).}
\end{figure}

Phonon data for $\alpha$-Sn (gray tin, \strb{} $A4$) was obtained by
Price {\em et al.}\cite{Price_PRB_3_1268_1971} at 90K, with a unit
cell volume of 68.1\AA$^3$.  Our closest volume was 68\AA$^3$, and we
compare our results with theirs along high-symmetry lines in the
Brillouin zone\cite{Setyawan_CMS_49_299_2010} in
Figure~\ref{fig:A4PBEph}.  The calculations are in good agreement with
experiment, especially in the longitudinal acoustic mode and the
optical modes.  This suggests that the harmonic approximation of APL
is adequate, and we do not need to look at anharmonic effects in tin.

\begin{figure}
  \includegraphics[width=0.5\textwidth]{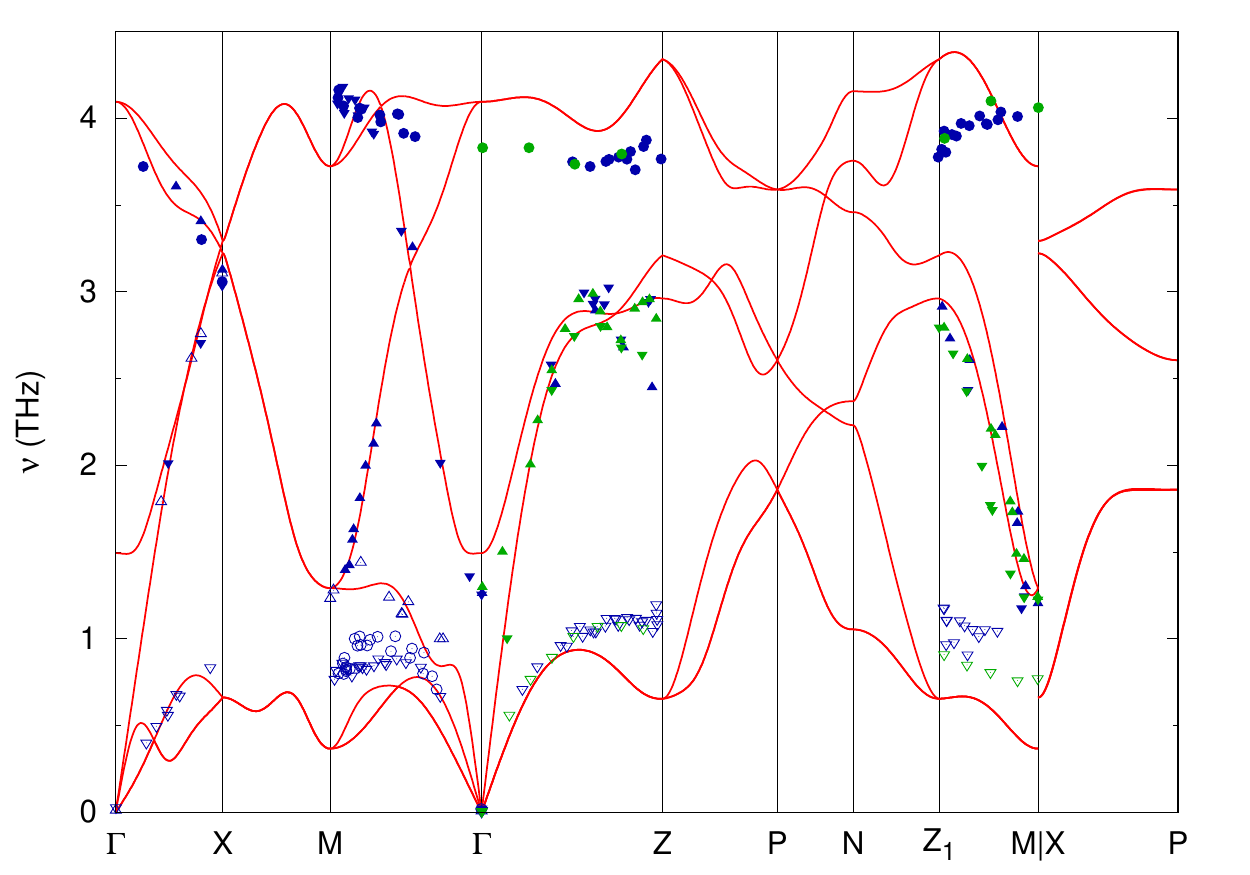}
  \caption{\label{fig:A5PBEph} Transverse (open symbols) and
    longitudinal (closed symbols) phonon frequencies for $\beta$-Sn
    measured at 300K by
    Price\cite{Price_Proc_Roy_Soc_Lon_A_300_25_1967} (blue symbols)
    and Rowe {\em et al.}\cite{Rowe_PRL_14_554_1965} at 296K (green
    symbols) compared to the PBE phonon spectra at the primitive cell
    volume of 54\AA$^3$ calculated using the APL module of AFLOW.  The
    high-symmetry paths through the body-centered tetragonal Brillouin
    zone are defined by Setyawan and
    Curtarolo.\cite{Setyawan_CMS_49_299_2010} Note that Price only
    determined the frequencies of one acoustic and one optic
    transverse branch along the $\Gamma-X$ ($\Delta$) line.  Data
    points were obtained from the references using the Engauge
    Digitizer.\cite{Mitchell_Engauge_2019}}
\end{figure}

Price\cite{Price_Proc_Roy_Soc_Lon_A_300_25_1967} and Rowe {\em et
  al.}\cite{Rowe_PRL_14_554_1965} measured the phonon spectrum of
$\beta$-Sn (white tin, \strb{} $A5$) at 300K and 296K, respectively.
At this temperature the primitive cell volume is 54.1\AA$^3$, so we
compare with our calculations at 54\AA$^3$.  The results are shown in
Figure~\ref{fig:A5PBEph}.  The agreement with experiment is not as
quite as good as in $\alpha$-Sn, but the overall agreement is
adequate.

\begin{figure}
  \includegraphics[width=0.5\textwidth]{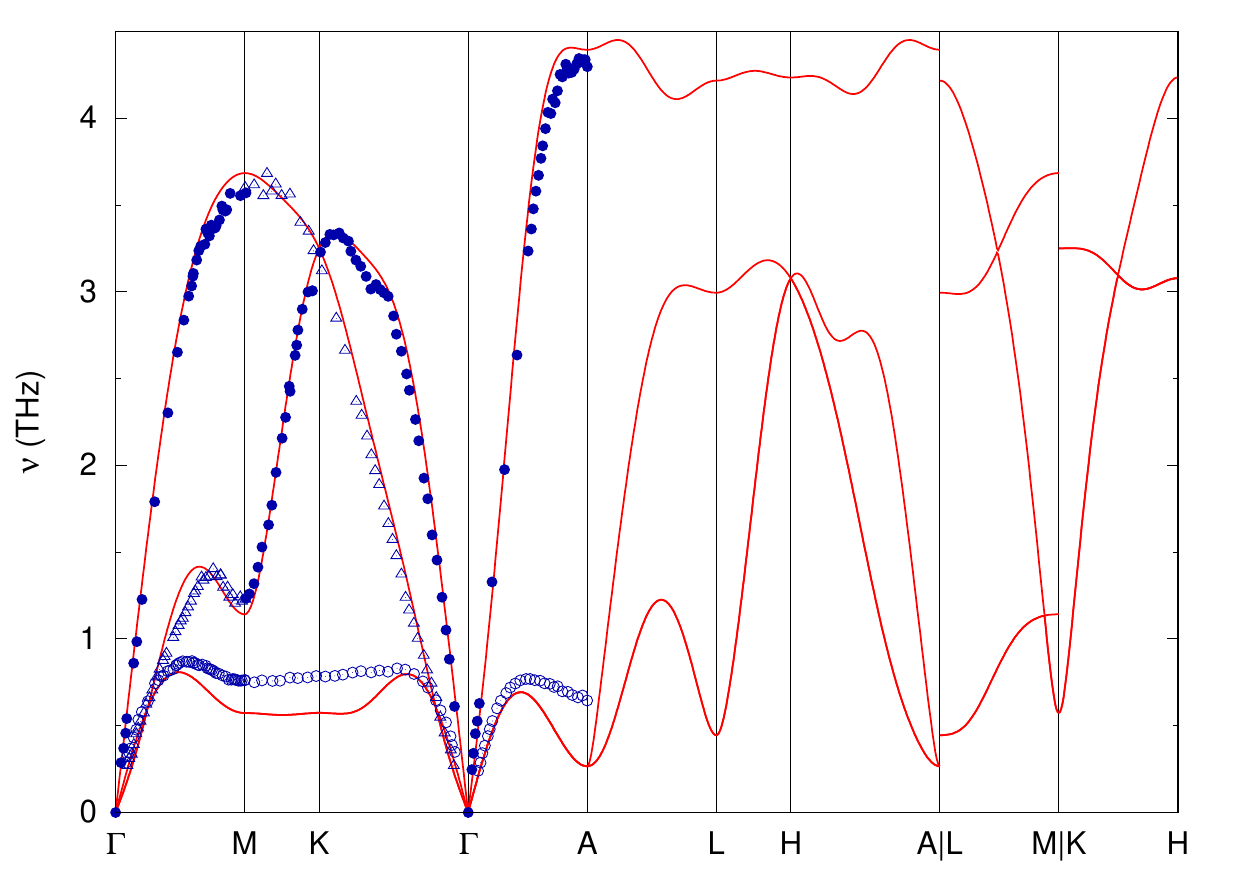}
  \caption{\label{fig:AfPBEph} Transverse (open symbols) and
    longitudinal (closed symbols) phonon frequencies for $\gamma$-Sn,
    Sn$_{0.8}$In$_{0.2}$ at room
    temperature,\cite{Ivanov_J_Phys_F_17_1925_1987} compared to the
    PBE phonon spectra at a primitive cell volume of 27\AA$^3$
    calculated using the APL module of AFLOW.  The high-symmetry paths
    through the simple hexagonal Brillouin zone are defined by
    Setyawan and Curtarolo.\cite{Setyawan_CMS_49_299_2010} Data points
    were obtained from the references using the Engauge
    Digitizer.\cite{Mitchell_Engauge_2019}}
\end{figure}

There are no samples of simple hexagonal (\strb{} $A_{f}$) tin, but
alloying with indium is known to stabilize this phase.  Ivanov {\em et
  al.}\cite{Ivanov_J_Phys_F_17_1925_1987} measured the phonon spectrum
of $\gamma$-Sn Sn$_{0.8}$In$_{0.2}$ at room temperature, where they
found the sample to have a primitive cell volume of 26.8\AA$^3$.  We
compare that to our calculations for pure simple hexagonal tin at
27\AA$^3$ in Figure~\ref{fig:AfPBEph}.  Here the agreement is
excellent.

As the figures show, the APL module can accurately predict the phonon
frequencies of tin in all three phases.  We can perform one more check
on the accuracy of the APL by determining the thermal expansion
coefficient for the $\alpha$- and $\beta$-Sn phases.  We could not
find data for the thermal expansion of $\gamma$-Sn, even as an alloy.

\subsection{\label{subsec:pbethermal} Thermal Expansion}

The APL module of AFLOW prints the vibrational free energy of at
intervals of 10K, so we can easily find the minimum free-energy volume
$V(T)$ over a large number of points.  To determine $\alpha(T)$ we fit
our data for $V(T)$ to the Pad\'{e} approximate
\begin{equation}
  \label{equ:vpade}
  V(T) = \frac{V_0 + a_1 T + a_2 T^2 + a_3 T^3}{1 + b_1 T + b_2 T^2} ~ ,
\end{equation}
which fits the expected linear behavior of $V(T)$ at high
temperatures.  From this we determine the linear expansion, averaged
over all directions, using (\ref{equ:alpha}).

\begin{figure}
    \includegraphics[width=0.5\textwidth]{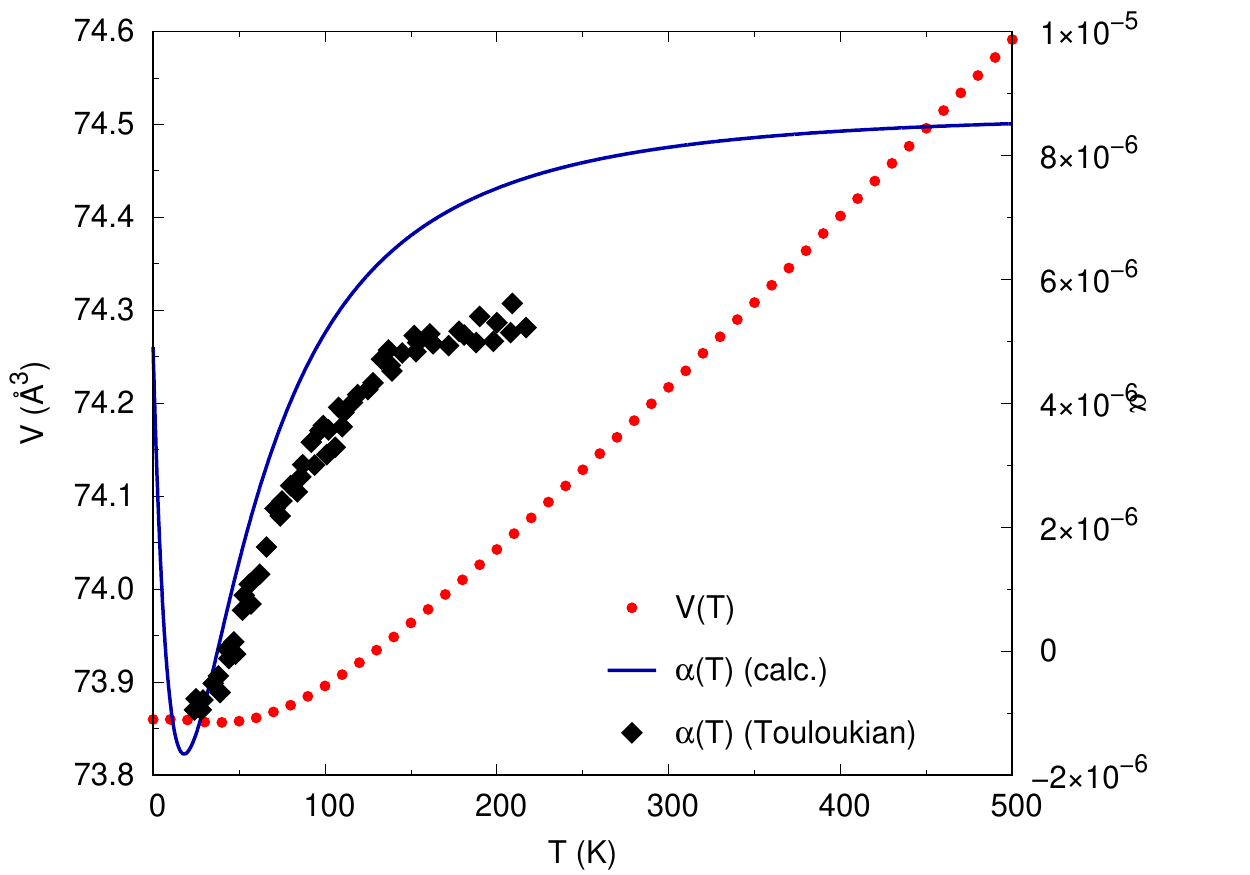}
  \caption{\label{fig:pbea4vt} The primitive-cell volume of
    $\alpha$-Sn as function of temperature (left axis, points), and
    linear expansion coefficient $\alpha$ (right axis, line)
    calculated by APL using the PBE density functional.  We also plot
    the experimental data found in Touloukian {\em et
      al.}\cite{Touloukian_Thermal_Properties_1975} (black
    diamonds).}
\end{figure}

The thermal expansion of $\alpha$-Sn is shown in
Figure~\ref{fig:pbea4vt}.  Above 200K $\alpha(T)$ is approximately
constant, tending toward a value of $8.5\times10^{-6}$.  Below 100K
our predictions are in excellent agreement with the experimental data
of Touloukian {\em et
  al.},\cite{Touloukian_Thermal_Properties_1975} although their
curve flattens out at a lower value than ours.

$\beta$-Sn has a tetragonal lattice, so the
lattice parameters $a(T)$ and $c(T)$ can have different thermal
expansion coefficients,
\begin{equation}
  \label{equ:tettherm}
  \alpha_{\parallel}(T) = \frac{1}{3a(T)} \frac{da}{dT}(T) ~ \mbox{and} ~
  \alpha_{\perp}(T) = \frac{1}{3c(T)} \frac{dc}{dT}(T) ~ ,
\end{equation}
where $\parallel$ and $\perp$ denote expansion in the $a,b$ plane and
along the $c$ axis, respectively.  The averaged thermal expansion is
then
\begin{equation}
  \label{equ:avtherm}
  \alpha(T) = \frac13 \left[ 2 \alpha_{\parallel}(T) +
  \alpha_{\perp}(T)\right] ~ .
\end{equation}

\begin{figure}
    \includegraphics[width=0.5\textwidth]{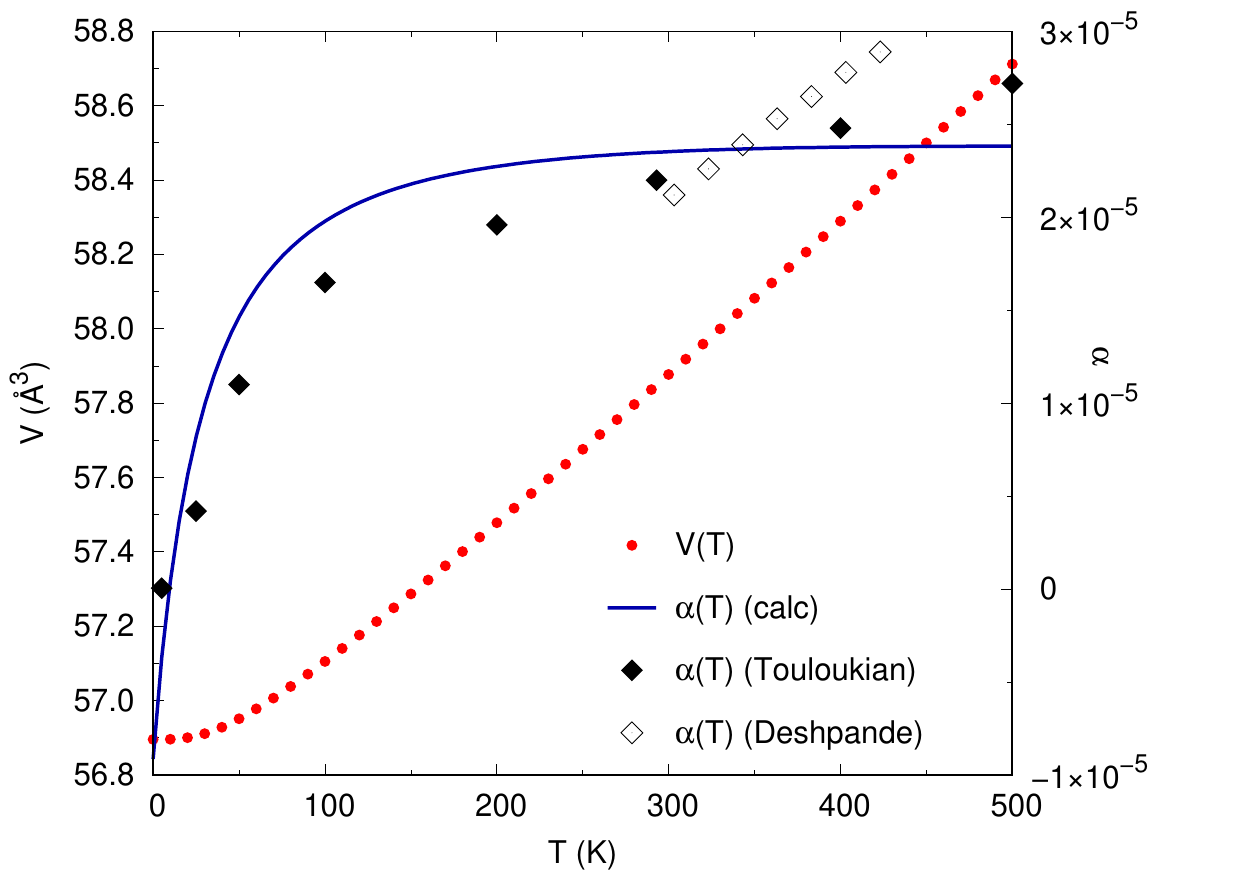}
  \caption{\label{fig:pbea5vt} The primitive-cell volume of $\beta$-Sn
    as function of temperature (left axis, points), and linear
    expansion coefficient $\alpha$ (right axis, line) calculated by
    APL using the PBE density functional.  We also plot the value of
    $\alpha$(T) found in Touloukian {\em et
      al.},\cite{Touloukian_Thermal_Properties_1975} (solid black
    diamonds) and measured by Deshpande and
    Sirdeshmukh\cite{Deshpande_Acta_Cryst_14_355_1961} (open black
    diamonds).}
\end{figure}

We compare predicted thermal expansion of $\beta$-Sn to the
experimental data cited in Touloukian {\em et
  al.},\cite{Touloukian_Thermal_Properties_1975} (solid black
diamonds) and measured by Deshpande and
Sirdeshmukh\cite{Deshpande_Acta_Cryst_14_355_1961} in Figure
\ref{fig:pbea5vt}.  We find good agreement with the experimental data,
but note that the experimental value of $\alpha$ is still increasing
at 500K, where our prediction is that it remains constant above 300K.

\begin{figure}
  \includegraphics[width=0.5\textwidth]{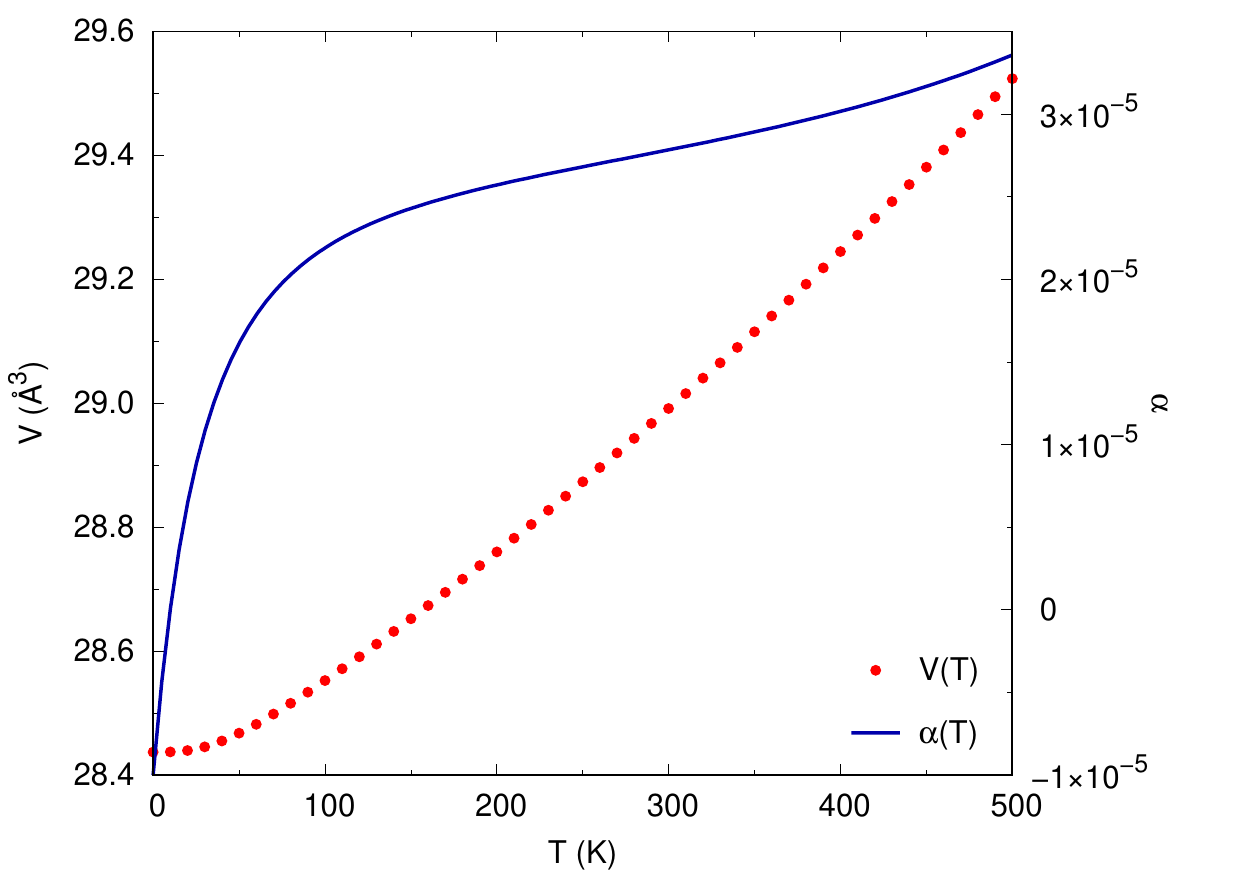}
  \caption{\label{fig:pbeafvt} The primitive-cell volume of
    $\gamma$-Sn as function of temperature (left axis, points), and
    linear expansion coefficient $\alpha$ (right axis, line)
    calculated by APL using the PBE density functional.}
\end{figure}

There is no experimental information about the thermal expansion of
$\gamma$-Sn or even Sn$_{0.8}$In$_{0.2}$, but we can determine the
thermal expansion parameter, which we plot in
Figure~\ref{fig:pbeafvt}.  We find that $\alpha(T)$ is still
increasing up to 500K, albeit not as rapidly as in experimental
$\beta$-Sn.  Around room temperature we find $\alpha \approx
2.8\times10^{-5}$, about 15\% larger than in $\beta$-Sn.

\subsection{\label{subsec:pbehull} Thermal Phase Transitions}

Now that we have shown that APL/AFLOW/VASP can compute both the phonon
spectra and thermal expansion of tin, at least while using the PBE
functional, we can turn to the question of phase stability: can we
predict the transition from $\alpha$- to $\beta$-Sn, and is there a
transition from $\beta$- to $\gamma$-Sn?

\begin{figure}
  \includegraphics[width=0.5\textwidth]{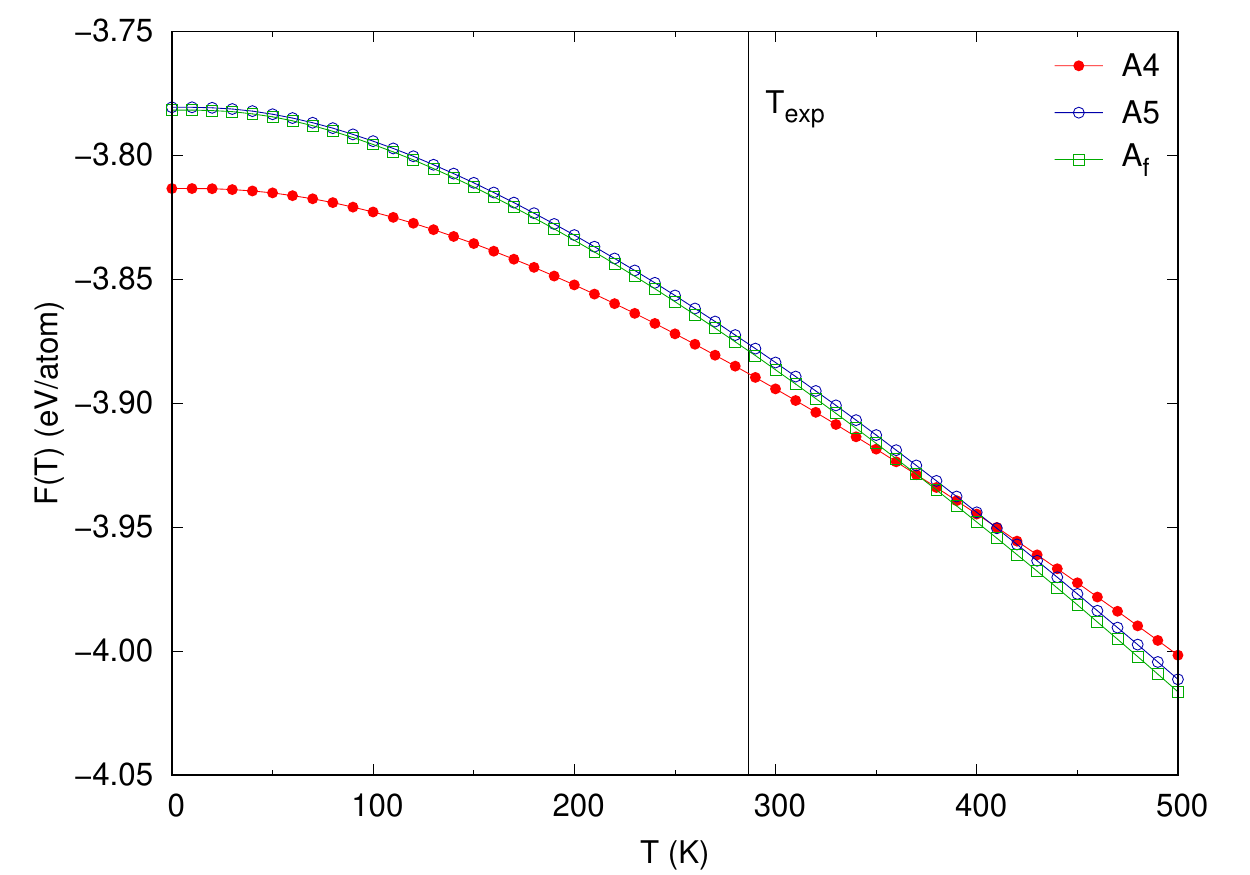}
  \caption{\label{fig:pbefree}Free energy as a function of temperature
    for $\alpha$-Sn (solid circles, \strb{} $A4$), $\beta$-Sn (open
    circles, \strb{} $A5$), and simple hexagonal $\gamma$-Sn (open
    squares, \strb{} $A_{f}$).  The vertical line shows the
    experimental $\alpha$-$\beta$ phase transition at 286K
    (13$^\circ$C).\cite{Cohen_Z_Phys_Chem_173_A_32_1935}}
\end{figure}

We have determined the free energy $F(T)$ for all three phases using
the PBE functional as outlined at the start of this section.  The
results are shown in Figure~\ref{fig:pbefree}.  We find that the
$\alpha$-$\beta$ phase transition occurs at 400K, some distance from
the experimental temperature of
286K,\cite{Cohen_Z_Phys_Chem_173_A_32_1935} but at least in the
correct range.  Unfortunately, the free energy of simple hexagonal
$\gamma$-Sn is always lower than $\beta$-Sn, and we predict an
$\alpha$-$\gamma$ transition at 370K.  This is not particularly
surprising given that static lattice calculations using PBE give
$U_{\gamma}$ lower than $U_{\beta}$ at volumes
(Figure~\ref{fig:pbeev}) and the phonon frequencies of $\beta$-Sn
(Figure~\ref{fig:A5PBEph}) and $\gamma$-Sn (Figure~\ref{fig:AfPBEph})
are spread over a comparable range of frequencies, suggesting that
their thermal properties should be similar.

\subsection{\label{subsec:thermscan} The SCAN Functional}

The SCAN functional correctly predicts the ordering
$U(\alpha\mbox{-Sn}) < U(\beta\mbox{-Sn}) < U(\gamma\mbox{-Sn})$
(Figure~\ref{fig:scanev}) so we expect that we will find the correct
ordering of thermal phase transitions.  The energy difference between
the $\alpha$- and $\beta$- phases is approximately 80\,meV/atom using
SCAN, compared to 40\,meV/atom using PBE, so that the predicted phase
transitions will be much higher than the transitions found using PBE,
which are already too large compared to experiment.  Given this, and
the cost of computing phonon frequencies using $\approx 250$ atom
supercells with a meta-GGA, we will not look at the phase transitions
using the SCAN functional at this time.

\begin{figure}
  \includegraphics[width=0.5\textwidth]{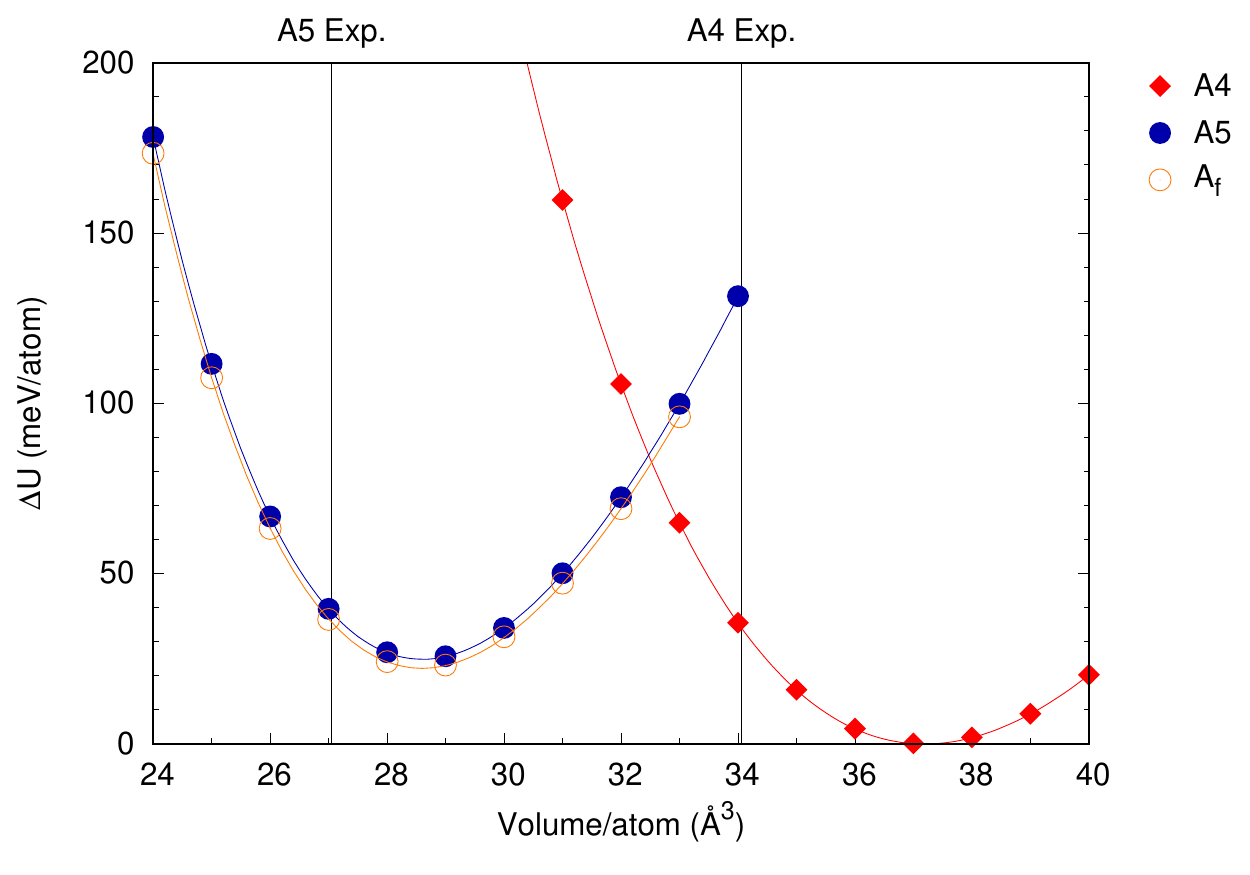}
  \caption{\label{fig:PBEU}Static lattice energy-volume curves for for
    $\alpha$- ($A4$), $\beta$- ($A5$), and $\gamma$-Sn ($A_{f}$),
    using LDA+U applied to the PBE functional with a shift $U_{p} =
    0.25$\,eV.  Since this is likely to predict a $\alpha \rightarrow
    \gamma$ transition we did not pursue further calculations.}
\end{figure}

\section{\label{sec:hubbard} Optimizing DFT via Coulomb Corrections}

Of course this work is not the first to study the $\alpha$-$\beta$
transition in tin, although it is the first to look in detail at the
$\gamma$-Sn problem.  Some time ago Pavone {\em
  al.}\cite{Pavone_PRB_57_10421_1998} found the $\alpha$-$\beta$ phase
transition at 311K, in reasonable agreement with experiment, but
probably influenced by errors in the LDA-based ultrasoft
pseudopotentials.\cite{Bachelet_PRB_26_4199_1982} Christensen and
Methfessel found that they could adjust the energy difference between
the $\alpha$- and $\beta$- phases by shifting the tin $4d$ orbitals
within a full-potential Linearized Muffin-Tin Orbital
scheme.\cite{Christensen_PRB_48_5797_1993} Legrain {\em
  et.\,al}\cite{Legrain_JCP_143_204701_2015} simply adjusted the
$\alpha$-$\beta$ energy difference until the curves in
Figure~\ref{fig:pbefree} crossed at the experimental transition
temperature, and Legrain and
Manzhos\cite{Legrain_AIPAdvances_6_045116_2016} showed that applying a
Hubbard U correction of 1\,eV to the tin $4s$ orbital changes the
energy difference from 40\,meV to 23\,meV, approximately what is
needed to lower the transition temperature to the experimental value.

\begin{figure}
  \includegraphics[width=0.5\textwidth]{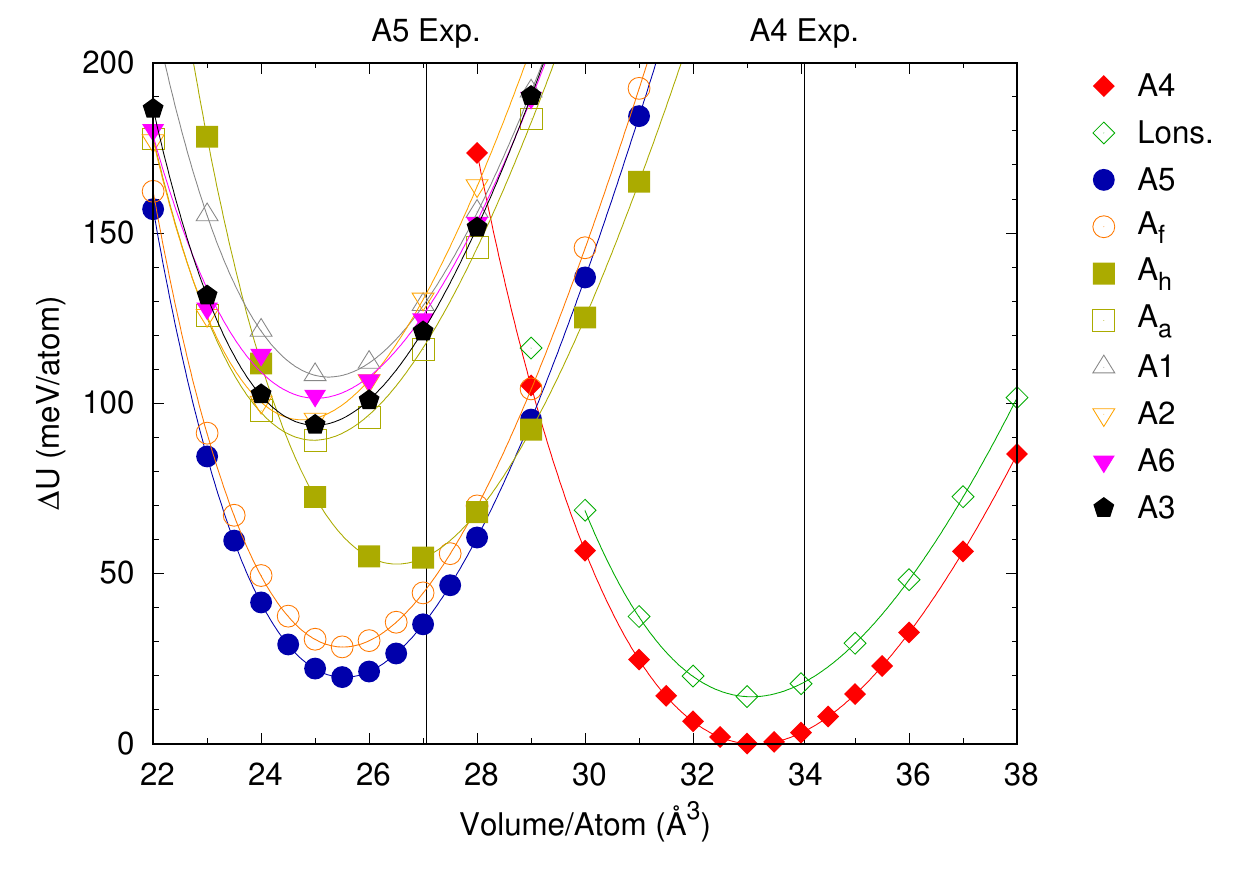}
  \caption{\label{fig:LDAU} Static lattice energy-volume curves for
    for the tin structures discussed in Section~\ref{sec:structures}
    as predicted by AFLOW/VASP using the LDA functional with a Hubbard
    correction $U_{p} = -0.8$\,eV.  The notation is identical to that in
    Figure~\ref{fig:ldaev}.}
\end{figure}

This last work led us to investigate the effect of a Hubbard U
correction on the $\alpha$-$\beta$ transition with
VASP.\cite{Bengone_PRB_62_16392_2000} VASP allows us to apply the
Hubbard U to a single orbital for each atomic species in the
calculation.  We chose to shift the $p$ orbitals using the LDSA+U
scheme of Liechtenstein.\cite{Liechtenstein_PRB_52_R5467_1995} This is
purely a variational parameter, as we do not have a physical reason
for doing this except the expectation that we can change the relative
energies of these systems.  We experimented with the choice of $U$ in
order to shift the $\alpha$-$\beta$ energy difference to approximately
20\,meV, which should give us a transition temperature on the order
of experiment.

We first tried this technique with the PBE functional, which would
seem to need the smallest correction.  If we take a value for $U_{p} =
0.25$\,eV, we lower the equilibrium $\alpha$-$\beta$ energy difference
to approximately 25\,meV/atom, as shown in Figure~\ref{fig:PBEU}.
This does not change the relative order of the phases, and so it will
still predict a phase transformation from $\alpha$-Sn to $\gamma$-Sn
rather than to $\beta$-Sn.

We then tried the same technique using the LDA functional.  If we
apply a shift $U_{p} = -0.8$\,eV, we get the results shown if
Figure~\ref{fig:LDAU}.  This is much more promising.  The
$\alpha$-$\beta$ energy difference is now 18\,meV, and the $\gamma$-Sn
phase is above the $\beta$-phase for all volumes studied.  In fact the
energy difference between $\beta$- and $\gamma$-Sn phase is large
enough so that the barrier between the the two phases vanishes.
Frozen-phonon calculations using FROZSL from the ISOTROPY software
Suite\cite{Stokes:FROZSL} show that the $L_3^-$ phonon has an
imaginary frequency.  This instability is exactly what is required to
transform the simple hexagonal structure to
$\beta$-Sn,\cite{Needs_PRB_30_5390_1984} and the frozen-phonon
supercells produced by FROZSL indeed relax to $\beta$-Sn.

\begin{figure}
    \includegraphics[width=0.5\textwidth]{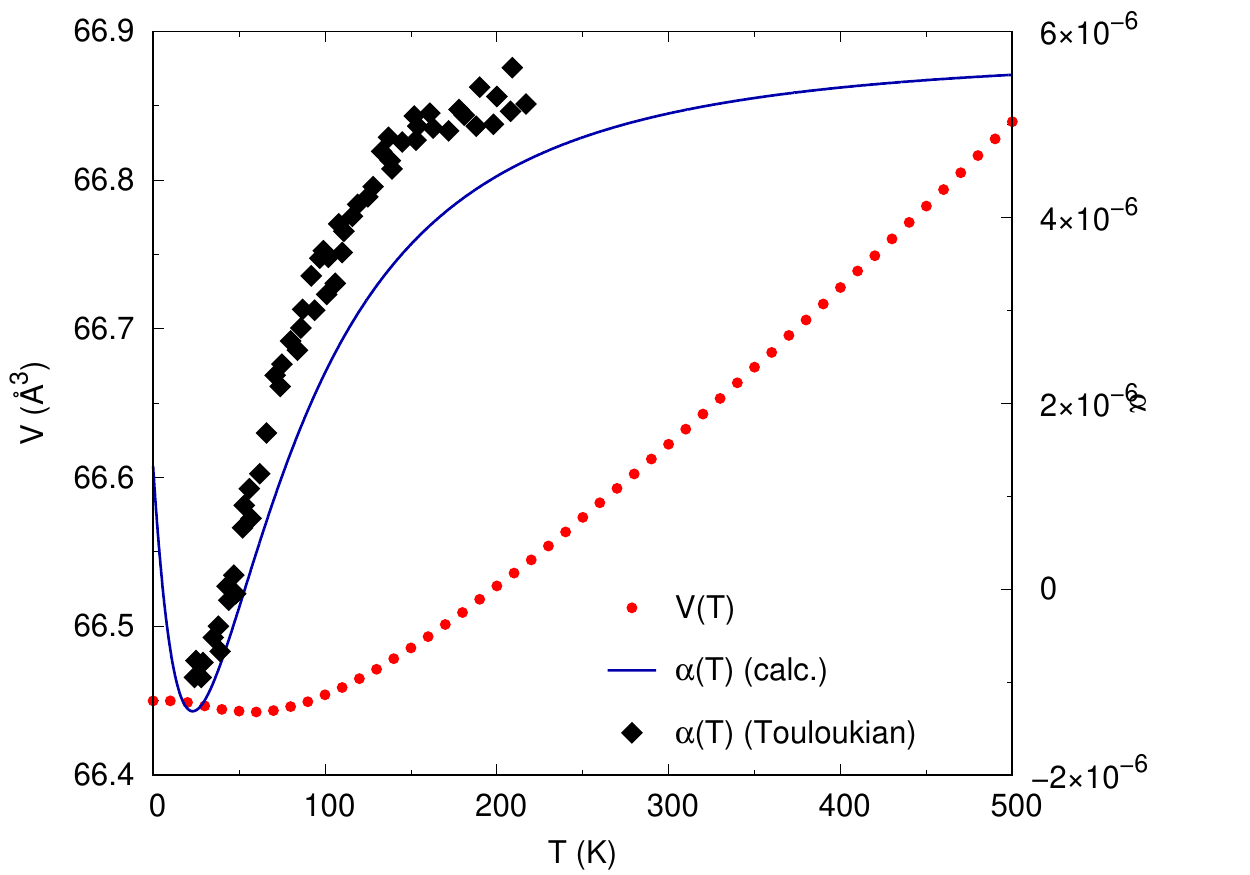}
  \caption{\label{fig:ldaua4vt} The primitive-cell volume of
    $\alpha$-Sn as function of temperature (left axis, points), and
    linear expansion coefficient $\alpha$ (right axis, line)
    calculated by APL using LDAU with $U_{p} = -0.8$\,eV.  We also
    plot the experimental data found in Touloukian {\em et
      al.}\cite{Touloukian_Thermal_Properties_1975} (black
    diamonds).}
\end{figure}

\begin{figure}
    \includegraphics[width=0.5\textwidth]{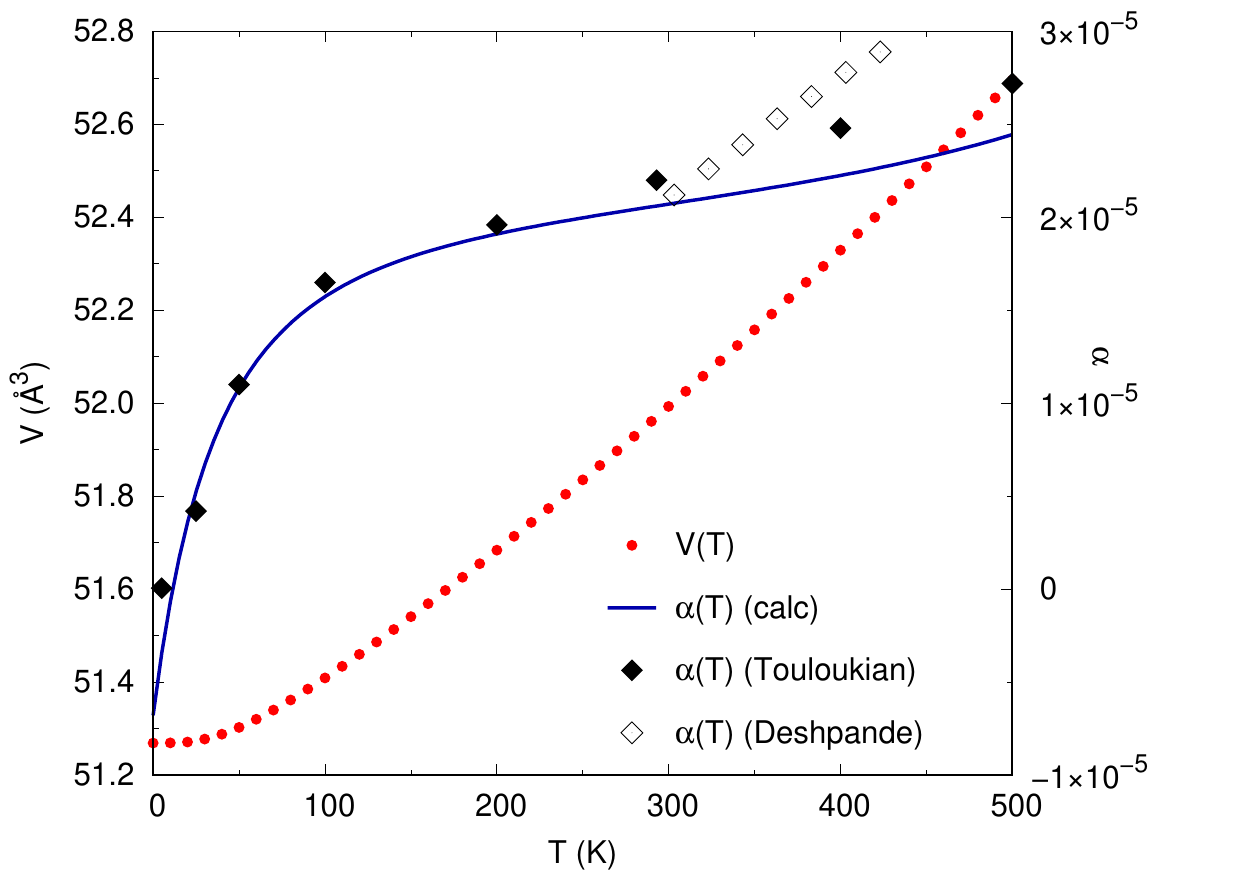}
  \caption{\label{fig:ldaua5vt} The primitive-cell volume of
    $\beta$-Sn as function of temperature (left axis, points), and
    linear expansion coefficient $\alpha$ (right axis, line)
    calculated by APL using the LDAU with $U_{p} = -0.8$\,eV.  We also
    plot the value of $\alpha$(T) found in Touloukian {\em et
      al.},\cite{Touloukian_Thermal_Properties_1975} (solid black
    diamonds) and measured by Deshpande and
    Sirdeshmukh\cite{Deshpande_Acta_Cryst_14_355_1961} (open black
    diamonds).}
\end{figure}

We can look at the thermal expansion in both $\alpha$-
(Figure~\ref{fig:ldaua4vt}) and $\beta$-Sn
(Figure~\ref{fig:ldaua5vt}).  In both cases the thermal expansion is
somewhat smaller than found using the PBE.  Our prediction for
$\alpha(T)$ for $\alpha$-Sn is slightly smaller than experiment, but
in $\beta$-Sn we find excellent agreement with experiment up to room
temperature.

\begin{figure}
  \includegraphics[width=0.5\textwidth]{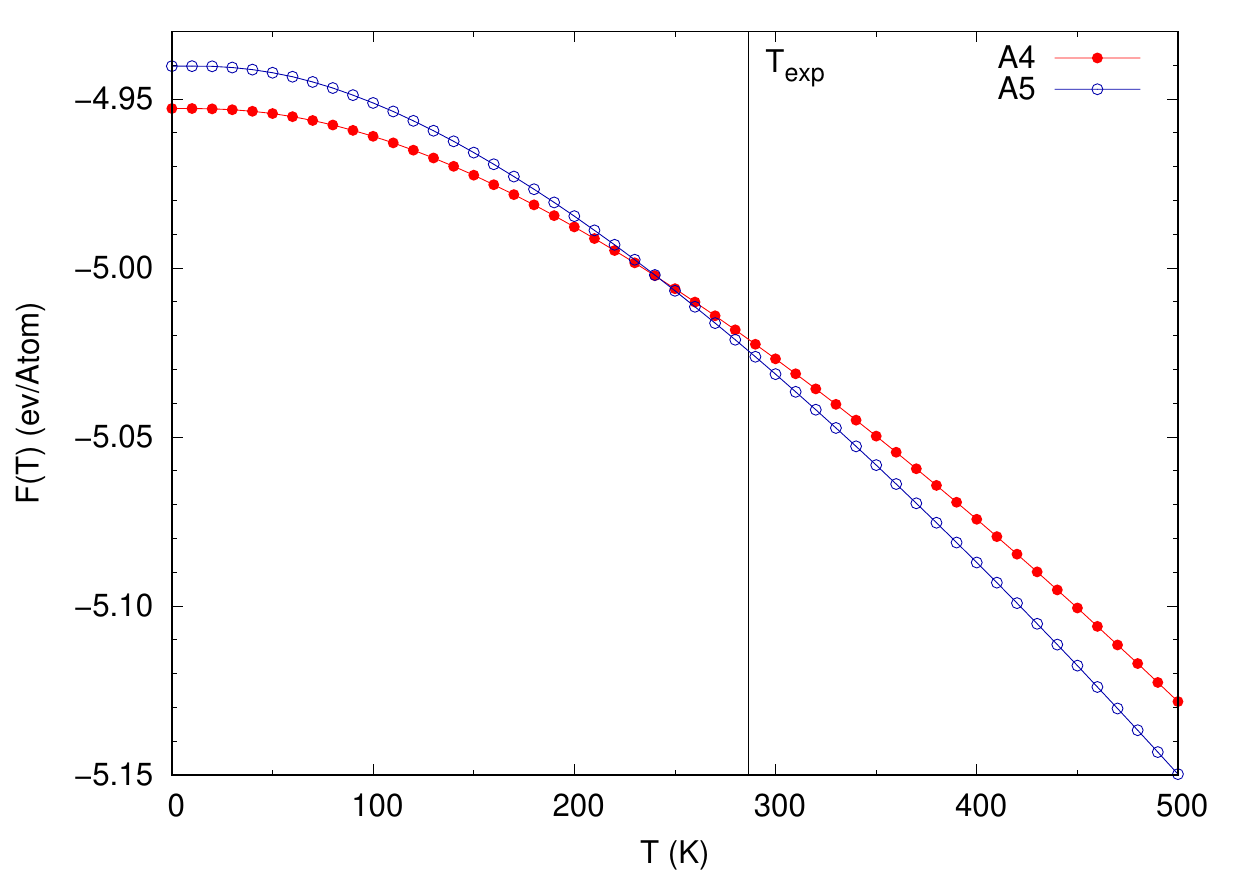}
  \caption{\label{fig:LDAUFree}Free energy as a function of
    temperature for $\alpha$-Sn (solid circles, \strb{} $A4$) and
    $\beta$-Sn (open circles, \strb{} $A5$), using the LDAU a Hubbard
    shift $U_{p} = -0.8$\,eV.  The vertical line shows the experimental
    $\alpha$-$\beta$ phase transition at 286K
    (13$^\circ$C).\cite{Cohen_Z_Phys_Chem_173_A_32_1935}}
\end{figure}

The thermal phase transition is again computed by the procedure
outlined in Section~\ref{sec:thermal}, but this time we using the LDA
and $U_{p} = -0.8$\,eV, obtaining the free energies shown in
Figure~\ref{fig:LDAUFree}.  Now the transition is predicted to occur
at 245K.  With further adjustment of $U_{p}$ we can obviously shift
the curves so that the transition is at the experimental value of
286K.

\begin{table*}
  \caption{\label{tab:converge} Effect of changing the kinetic energy
    cutoff and k-point mesh density on the equilibrium properties of
    $\alpha$-, $\beta$-, and $\gamma$-Sn, for the LDA, PBE, and SCAN
    functionals used in VASP.  The left-hand column under each
    functional uses the AFLOW default kinetic energy cutoff (ENMAX) and
    k-point mesh (KMESH), while the right-hand column uses a higher
    energy cutoff and denser k-point mesh, as shown.  Equilibrium
    volume ($V_0$) and $c/a$ ratios are computed from the final
    configuration found by VASP.  The volume $V_0$ is the volume of
    the primitive cell for each atom, but the energy differences are
    per atom.}
  \begin{tabular}{c|c|c|c|c|c|c}
    \hline
    & \multicolumn{2}{c|}{LDA} & \multicolumn{2}{c|}{PBE} &
    \multicolumn{2}{c}{SCAN} \\
    \hline
    ENMAX & 144.6 & 350.0 & 144.6 & 350.0 & 337.5 & 725.0 \\
    \hline
    \multicolumn{7}{c}{$\alpha$-Sn Equilibrium Configuration} \\
    \hline
    KMESH & $16\times16\times16$ & $24\times24\times24$ &
    $16\times16\times16$ & $24\times24\times24$ & $16\times16\times16$
    & $24\times24\times24$ \\
    $V_0$ (\AA$^3$) & 67.98413 & 67.95582 & 73.62472 & 73.58994 & 69.67733
    & 70.64511 \\
    \hline
    \multicolumn{7}{c}{$\beta$-Sn Equilibrium Configuration} \\
    \hline
    KMESH & $24\times24\times24$ & $32\times32\times32$ &
    $24\times24\times24$ & $32\times32\times32$ & $24\times24\times24$
    & $32\times32\times32$ \\
    $V_0$ (\AA$^3$) & 52.36742 & 52.39377 & 56.64443 & 56.69880 & 54.39482
    & 55.08161 \\
    $c/a$ & 0.54174 & 0.54117 & 0.54186 & 0.54175 & 0.54111 & 0.53850 \\
    \hline
    \multicolumn{7}{c}{$\gamma$-Sn Equilibrium Configuration} \\
    \hline
    KMESH & $21\times21\times19$ & $36\times36\times36$ &
    $21\times21\times19$ & $36\times36\times36$ & $21\times21\times19$
    & $36\times36\times36$ \\
    $V_0$ (\AA$^3$) & 26.16382 & 26.19763 & 28.33760 & 28.35201 & 27.13348
    & 27.48595 \\
    $c/a$ & 0.93782 & 0.93825 & 0.93969 & 0.93982 & 0.93832 & 0.93818 \\
    \hline
    \multicolumn{7}{c}{Equilibrium Energy Differences (eV/atom)} \\
    \hline
    U($\beta$-Sn) - U($\alpha$-Sn) & -0.02309 & -0.02324 & 0.03961 &
    0.03856 & 0.08123 & 0.07261 \\
    U($\gamma$-Sn) - U($\alpha$-Sn) & -0.02101 & -0.02041 & 0.03913 &
    0.03840 & 0.08573 & 0.08001 \\
    U($\gamma$-Sn) - U($\beta$-Sn) & 0.00208 & 0.00284 & -0.00048 &
    -0.00015 & 0.00450 & 0.00740 \\
    \hline
  \end{tabular}
\end{table*}

\section{\label{sec:convergence} High-Throughput Calculations and
  Convergence}

By their nature, high-throughput calculations rely on a set of
standard assumptions, in particular that the basis set size (kinetic
energy cutoff in a plane-wave code) and k-point mesh can be fixed
without regard to the crystal structure being studied.  For example,
by default AFLOW sets the kinetic energy cutoff (ENMAX in VASP) to
140\% of the minimum value recommended by VASP and the k-point mesh to
give a minimum of 8000 k-points per reciprocal atom in the Brillouin
zone,\cite{Calderon_Comp_Mat_Sci_108A_233_2015} equivalent to a $20
\times 20 \times 20$ k-point mesh for a cubic system.  While these
standard values are usually sufficient, they may lead to errors when
energy differences between phases are small.

We tested the reliability of the default energy cutoff and k-point
size for tin by performing two sets of calculations to find the
minimum energy configuration for the $\alpha$-, $\beta$-, and
$\gamma$-Sn phases.  The first set used the AFLOW default values for
both quantities.  The second set approximately doubled the energy
cutoff, and increased the k-point density in the Brillouin zone.  The
results are shown in Table~\ref{tab:converge}.  There is little
difference between the two calculations' equilibrium volumes and
$c/a$.  There is a larger discrepancy in the energy differences.  In
the LDA the change can be as much as 0.6\,meV, while in SCAN the
change is as much as 3\,meV.  Fortunately this does not change the
overall conclusions of the preceding sections.

\begin{figure}
    \includegraphics[width=8cm]{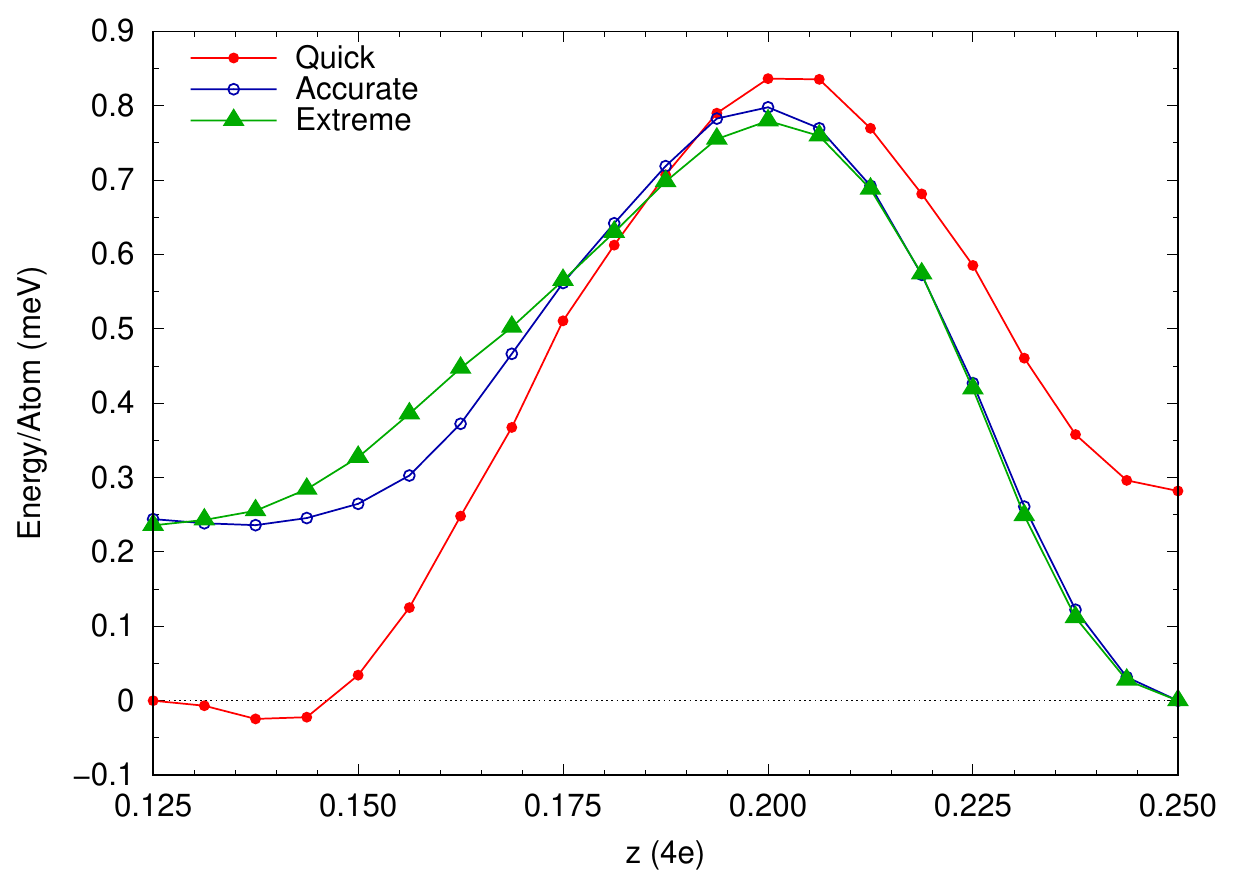}
  \caption{\label{fig:barrier} (color online) Energy barrier for the
    path from $\beta$-Sn to $\gamma$-Sn from Needs and
    Martin\cite{Needs_PRB_30_5390_1984} (\ref{equ:needsmartin}).  The
    closed-circle red symbols are from the ``quick'' calculations, the
    open blue symbols are from the ``accurate'' calculations, and the
    green triangles from the ``extreme'' calculations described in the
    text.  The energy differences between the end-point phases shown
    here differ from those in Table~\ref{tab:equprop} because we have
    not reached convergence in the k-point mesh, and the unit cells
    and k-point meshes used to describe the $\gamma$-Sn phase differ
    in the two cases.}
\end{figure}

For an even more extreme example of the importance of energy cutoffs
and k-point meshes, consider the energy barrier along the Needs-Martin
path (\ref{equ:needsmartin}): compute the barrier between $\beta$- and
$\gamma$-Sn by fixing $z$ and allowing VASP to relax the orthorhombic
cell's lattice constants.  Starting positions for the orthorhombic
lattice for each $z$ value are linearly interpolated from the end
points.  We did this calculation with the PBE functional with
``quick'' settings of ENMAX = 144.6\,eV and KMESH=$32\times32\times32$
(4504 k-points in the irreducible Brillouin zone); with the
``accurate'' settings, ENMAX = 350\,eV and KMESH=$36\times36\times36$
(6346 k-points); and an ``extreme'' k-point mesh, ENMAX = 350\,eV and
KMESH=$48\times48\times48$ (14,725 k-points).  The results are shown
in Figure~\ref{fig:barrier}.

Here we find that the lowest energy structure shifts from the quick to
the accurate cases, and that the energy difference between the
structures does not agree with what we found in
Table~\ref{tab:converge} even in the ``accurate'' or ``extreme''
cases.  This is because the k-point meshes we use, while adequate for
$\beta$-Sn structure, are far from ideal for the hexagonal lattice of
the $\gamma$-Sn structure.  The smaller k-point meshes do not predict
the correct behavior for small distortions away from the $\beta$-Sn
structure, predicting that the $\beta$-Sn phase is not even metastable
within the PBE.  This instability would correspond to an imaginary
transverse acoustic phonon near $\Gamma$ in Figure~\ref{fig:A5PBEph}.
The stability of the $\beta$-Sn phase can only be confirmed with the
denser k-point mesh.

\section{\label{sec:conclude} Summary}

We used first-principles density functional calculations to study the
$\alpha$-Sn $\rightarrow$ $\beta$-Sn phase transition and the possible
transition to a hexagonal $\gamma$-Sn phase using a variety of density
functionals available in VASP within the AFLOW high-throughput
framework.  The SCAN functional is the only one which predicts the
correct sequence, $U(\alpha\mbox{-Sn}) < U(\beta\mbox{-Sn}) <
U(\gamma\mbox{-Sn})$, and there the energy difference
$U(\beta\mbox{-Sn}) - U(\alpha\mbox{-Sn})$ is much too large to
account for the low temperature of the phase transition.  The only
other functional which predicts the $\beta$-Sn phase to be above
$\alpha$-Sn is PBE, but it orders the functions so that
$U(\alpha\mbox{-Sn}) < U(\gamma\mbox{-Sn}) < U(\beta\mbox{-Sn})$.  If
we ignore the $\gamma$-Sn phase we get the $\alpha$ $\rightarrow$
$\beta$ transition at 400K, some 120K above experiment.  Of the
remaining functionals, the LDA and other GGA functionals all predict
the $\beta$ phase to be the ground state of tin, and the non-SCAN
meta-GGAs predict that the hexagonal close-packed phase is the ground
state, and indeed overbind all of the close-packed phases.

We used a Hubbard $U$ Coulomb correction as a fitting parameter, and
found that we can adjust the value of $U_{p}$ so that we can get the
$\alpha$-$\beta$ energy difference needed to get the phase transition
near 286K.  Applying a Hubbard $U$ to the PBE functional still leaves
the $\gamma$-Sn phase lower in energy than $\beta$-Sn, but a
correction to the LDA will get us to the correct transition
temperature, and the $\gamma$-Sn phase is unstable and relaxes to
$\beta$-Sn.  Whether this last finding is correct or not depends on
the existence of the hexagonal or near-hexagonal $\gamma$-Sn phase
described by Kubiak.\cite{Kubiak_JLCM_116_307_1986}

We have also looked to see if our predictions change when we use
basis-set sizes and k-point meshes larger than the AFLOW default.  We
find that our overall predictions do not change, but care must be
taken if very accurate results are desired.

In conclusion, none of the tested density functionals actually predict
the proper behavior of tin, making this system an ideal test for the
evaluation of new functionals.

\begin{acknowledgments}
The authors thank
Ohad Levy and Cormac Toher
for valuable discussions.
 This work has been supported by ONR Grants N00014-20-1-2525 and N00014-20-1-2200.
  R.F. acknowledges support from the Alexander von Humboldt foundation
  under the Feodor Lynen research fellowship.  Some calculations were
  performed on the United States Naval Academy Cray, Grace.
\end{acknowledgments}

\newcommand{\Ozolins}{Ozoli{\c{n}}{\v{s}}}


\begin{thebibliography}{10}
\expandafter\ifx\csname urlstyle\endcsname\relax
  \providecommand{\doi}[1]{doi:\discretionary{}{}{}#1}\else
  \providecommand{\doi}{doi:\discretionary{}{}{}\begingroup
  \urlstyle{rm}\Url}\fi
\providecommand{\selectlanguage}[1]{\relax}
\providecommand{\bibAnnoteFile}[1]{%
  \IfFileExists{#1}{\begin{quotation}\noindent\textsc{Key:} #1\\
  \textsc{Annotation:}\ \input{#1}\end{quotation}}{}}
\providecommand{\bibAnnote}[2]{%
  \begin{quotation}\noindent\textsc{Key:} #1\\
  \textsc{Annotation:}\ #2\end{quotation}}

\bibitem{Mehl_Comp_Mat_Sci_136_S1_2017}
M.~J. Mehl, D.~Hicks, C.~Toher, O.~Levy, R.~M. Hanson, G.~Hart, and
  S.~Curtarolo, \emph{The AFLOW Library of Crystallographic Prototypes: Part
  1}, Computational Materials Science \textbf{136}, S1--S828 (2017).
\input{Mehl_Comp_Mat_Sci_136_S1_2017}

\bibitem{Cohen_Z_Phys_Chem_173_A_32_1935}
E.~Cohen and A.~K. W.~A. van Lieshout, \emph{Physikalisch-chemische Stadien am
  Zinn. X. Die Umwandlungstemperatur graues Zinn $\rightleftharpoons$ weisses
  Zinn}, Zeitschrift f\"{u}r Physikalische Chemie A \textbf{173}, 32--34
  (1935).
\input{Cohen_Z_Phys_Chem_173_A_32_1935}

\bibitem{Cornelius_Microelec_Rel_79_175_2017}
B.~Cornelius, S.~Treivish, Y.~Rosenthal, and M.~Pecht, \emph{The phenomenon of
  tin pest: A review}, Microelectronics Reliability \textbf{79}, 175--192
  (2017).
\input{Cornelius_Microelec_Rel_79_175_2017}

\bibitem{Fritzsche_Berichte_Deutsche_Chem_Gesell_1869}
J.~Fritzsche, \emph{Ueber eigenth\"{u}lich modificirtes Zinn}, Berichte der
  deutschen chemischen Gesellschaft pp. 112--113 (1869).
\input{Fritzsche_Berichte_Deutsche_Chem_Gesell_1869}

\bibitem{Plumbridge_JMS_Mater_Elec_18_307_2007}
W.~J. Plumbridge, \emph{Tin pest issues in lead-free electronic solders},
  Journal of Materials Science: Materials in Electronics \textbf{18}, 307--318
  (2007).
\input{Plumbridge_JMS_Mater_Elec_18_307_2007}

\bibitem{LeCouteur_Napoleons_Buttons_2004}
P.~L. Couteur and J.~Burreson, \emph{Napoleon's Buttons: How 17 Molecules
  Changed History} (Tarcher Perigee, New York, 2004).
\input{LeCouteur_Napoleons_Buttons_2004}

\bibitem{Pfister_arxiv_1204_1443_2012}
R.~Pfister and P.~Pugnat, \emph{Tin Pest: A Forgotten Issue in the Field of
  Applied Superconductivity?} (2012). ArXiv:1204.1443 [cond-mat.supr-con].
\input{Pfister_arxiv_1204_1443_2012}

\bibitem{Kubiak_JLCM_116_307_1986}
R.~Kubiak, \emph{Evidence for the existence of the $\gamma$ form of tin},
  Journal of the Less Common Metals \textbf{116}, 307--311 (1986).
\input{Kubiak_JLCM_116_307_1986}

\bibitem{Kane_Acta_Met_14_605_1966}
R.~H. Kane, B.~C. Giessen, and N.~J. Grant, \emph{New metastable phases in
  binary tin alloy systems}, Acta Metallurgica \textbf{14}, 605--609 (1966).
\input{Kane_Acta_Met_14_605_1966}

\bibitem{Ivanov_J_Phys_F_17_1925_1987}
A.~S. Ivanov, A.~Y. Rumiantsev, B.~Dorner, N.~L. Mitrofanov, and V.~V.
  Pushkarev, \emph{Lattice dynamics and electron-phonon interaction in
  $\gamma$-tin}, Journal of Physics F: Metal Physics \textbf{17}, 1925--1934
  (1987).
\input{Ivanov_J_Phys_F_17_1925_1987}

\bibitem{Ivanov_Physica_B_174_79_1991}
A.~S. Ivanov, A.~Y. Rumiantsev, N.~L. Mitrofanov, and M.~Alba,
  \emph{Low-frequency lattice dynamics of $\gamma$-tin}, Physica B: Condensed
  Matter \textbf{174}, 79--82 (1974).
\input{Ivanov_Physica_B_174_79_1991}

\bibitem{Parthe_Gmelin_Handbook_1993}
E.~Parth\'{e}, L.~Gelato, B.~Chabot, M.~Penso, K.~Cenzula, and
  R.~Gladyshevskii, \emph{Standardized Data and Crystal Chemical
  Characterization of Inorganic Structure Types}, \emph{Gmelin Handbook of
  Inorganic and Organometallic Chemistry}, vol.~2 (Springer-Verlag, Berlin,
  Heidelberg, 1993), 8 edn.
\input{Parthe_Gmelin_Handbook_1993}

\bibitem{Pavone_PRB_57_10421_1998}
P.~Pavone, S.~Baroni, and S.~de~Gironcoli, \emph{$\alpha\leftrightarrow\beta$
  phase transition in tin: A theoretical study based on density-functional
  perturbation theory}, Physical Review B \textbf{1998}, 10421--10423 (57).
\input{Pavone_PRB_57_10421_1998}

\bibitem{Houben_PRB_100_075408_2019}
K.~Houben, J.~K. Jochum, D.~P. Lozano, M.~Bisht, E.~Men\'{e}ndez, D.~G. Merkel,
  R.~R\"{u}ffer, A.~I. Chumakov, S.~Roelants, B.~Partoens, M.~V.
  Milo\v{s}evi\'{c}, F.~M. Peeters, S.~Couet, A.~Vantomme, K.~Temst, and
  M.~J.~V. Bael, \emph{{\em In situ} study of the $\alpha$-Sn to $\beta$-Sn
  phase transition in low-dimensional systems: Phonon behavior and
  thermodynamic properties}, Physical Review B \textbf{100}, 075408 (2019).
\input{Houben_PRB_100_075408_2019}

\bibitem{Ihm_PRB_23_1576_1981}
J.~Ihm and M.~L. Cohen, \emph{Equilibrium properties and the phase transition
  of grey and white tin}, Physical Review B \textbf{23}, 1576--1579 (1981).
\input{Ihm_PRB_23_1576_1981}

\bibitem{Na_J_Korean_Phys_Soc_54_494_2010}
S.-H. Na and C.-H. Park, \emph{First-Principles Study of the Structural Phase
  Transition in Sn}, Journal of the Korean Physical Society \textbf{56},
  494--497 (2010).
\input{Na_J_Korean_Phys_Soc_54_494_2010}

\bibitem{Legrain_JCP_143_204701_2015}
F.~Legrain, O.~I. Malyi, C.~Persson, and S.~Manzhos, \emph{Comparison of alpha
  and beta tin for lithium, sodium, and magnesium storage: An {\em ab initio}
  study including phonon contributions}, Journal of Chemical Physics
  \textbf{143}, 204704 (2015).
\input{Legrain_JCP_143_204701_2015}

\bibitem{Legrain_AIPAdvances_6_045116_2016}
F.~Legrain and S.~Manzhos, \emph{Understanding the difference in cohesive
  energies between alpha and beta tin in DFT calculations}, AIP Advances
  \textbf{6}, 045116 (2016).
\input{Legrain_AIPAdvances_6_045116_2016}

\bibitem{Curtarolo_Comp_Mat_Sci_58_218_2012}
S.~Curtarolo, W.~Setyawan, G.~L.~W. Hart, M.~Jahn\'{a}tek, R.~V. Chepulskii,
  R.~H. Taylor, S.~Wang, J.~Xue, K.~Yang, O.~Levy, M.~J. Mehl, H.~T. Stokes,
  D.~O. Demchenko, and D.~Morgan, \emph{AFLOW: An automatic framework for
  high-throughput materials discovery}, Computational Materials Science
  \textbf{58}, 218--226 (2012).
\input{Curtarolo_Comp_Mat_Sci_58_218_2012}

\bibitem{Toher_AFLOW_Handbook_Mat_Model_1_2018}
C.~Toher, C.~Oses, D.~Hicks, E.~Gossett, F.~Rose, P.~Nath, D.~Usanmaz, D.~C.
  Ford, E.~Perim, C.~E. Calderon, J.~J. Plata, Y.~Lederer, M.~Jahnátek,
  W.~Setyawan, S.~Wang, J.~Xue, K.~Rasch, R.~V. Chepulskii, R.~H. Taylor,
  G.~Gomez, H.~Shi, A.~R. Supka, R.~A. R.~A. Orabi, P.~Gopal, F.~T. Cersoli,
  L.~Liyanage, H.~Wang, I.~Siloi, L.~A. Agapito, C.~Nyshadham, G.~L.~W. Hart,
  J.~Carrete, F.~Legrain, N.~Mingo, E.~Zurek, O.~Isayev, A.~Tropsha,
  S.~Sanvito, R.~M. Hanson, I.~Takeuchi, M.~J. Mehl, A.~N. Kolmogorov, K.~Yang,
  P.~D'Amico, A.~Calzolari, M.~Costa, R.~D. Gennaro, M.~B. Nardelli,
  M.~Fornari, O.~Levy, and S.~Curtarolo, \emph{The AFLOW Fleet for Materials
  Discovery}, in \emph{Handbook of Materials Modeling}, edited by W.~Andreoni
  and S.~Yip (Springer International Publishing, Cham, Switzerland, 2018), pp.
  1--28.
\input{Toher_AFLOW_Handbook_Mat_Model_1_2018}

\bibitem{Oses_MRS_Bull_43_670_2018}
C.~Oses, C.~Toher, and S.~Curtarolo, \emph{Data-driven design of inorganic
  materials with the Automatic Flow Framework for Materials Discovery}, MRS
  Bulletin \textbf{43}, 670--675 (2018).
\input{Oses_MRS_Bull_43_670_2018}

\bibitem{curtarolo:art113}
C.~Nyshadham, C.~Oses, J.~E. Hansen, I.~Takeuchi, S.~Curtarolo, and G.~L.~W.
  Hart, \emph{A computational high-throughput search for new ternary
  superalloys}, Acta\ Mater. \textbf{122}, 438--447 (2017).
\input{curtarolo:art113}

\bibitem{curtarolo:art67}
M.~Jahn{\'{a}}tek, O.~Levy, G.~L.~W. Hart, L.~J. Nelson, R.~V. Chepulskii,
  J.~Xue, and S.~Curtarolo, \emph{Ordered phases in ruthenium binary alloys
  from high-throughput first-principles calculations}, Phys.\ Rev.\ B
  \textbf{84}, 214110 (2011).
\input{curtarolo:art67}

\bibitem{curtarolo:art139}
Y.~Lederer, C.~Toher, K.~S. Vecchio, and S.~Curtarolo, \emph{The search for
  high entropy alloys: A high-throughput \textit{ab-initio} approach}, Acta\
  Mater. \textbf{159}, 364--383 (2018).
\input{curtarolo:art139}

\bibitem{Kresse_PRB_47_558_1993}
G.~Kresse and J.~Hafner, \emph{Ab initio molecular dynamics for liquid metals},
  Physical Review B \textbf{47}, 558--561 (1993).
\input{Kresse_PRB_47_558_1993}

\bibitem{Kresse_PRB_49_14251_1994}
G.~Kresse and J.~Hafner, \emph{{\em Ab initio} molecular-dynamics simulation of
  the liquid-metal-amorphous-semiconductor transition in germanium}, Physical
  Review B \textbf{49}, 14251--14269 (1994).
\input{Kresse_PRB_49_14251_1994}

\bibitem{Kresse_Comp_Mat_Sci_6_15_1996}
G.~Kresse and J.~Furthm\"{u}ller, \emph{Efficiency of {\em ab-initio} total
  energy calculations for metals and semiconductors using a plane-wave basis
  set}, Computational Materials Science \textbf{6}, 15--50 (1996).
\input{Kresse_Comp_Mat_Sci_6_15_1996}

\bibitem{Kresse_PRB_54_11169_1996}
G.~Kresse and J.~Furthm\"{u}ller, \emph{Efficient iterative schemes for {\em ab
  initio} total-energy calculations using a plane-wave basis set}, Physical
  Review B \textbf{54}, 11169--11186 (1996).
\input{Kresse_PRB_54_11169_1996}

\bibitem{curtarolo:art65}
S.~Curtarolo, W.~Setyawan, G.~L.~W. Hart, M.~Jahn{\'{a}}tek, R.~V. Chepulskii,
  R.~H. Taylor, S.~Wang, J.~Xue, K.~Yang, O.~Levy, M.~J. Mehl, H.~T. Stokes,
  D.~O. Demchenko, and D.~Morgan, \emph{{AFLOW}: An automatic framework for
  high-throughput materials discovery}, Comput.\ Mater.\ Sci. \textbf{58},
  218--226 (2012).
\input{curtarolo:art65}

\bibitem{curtarolo:art125}
J.~J. Plata, P.~Nath, D.~Usanmaz, J.~Carrete, C.~Toher, M.~{de Jong}, M.~D.
  Asta, M.~Fornari, M.~{Buongiorno Nardelli}, and S.~Curtarolo, \emph{An
  efficient and accurate framework for calculating lattice thermal conductivity
  of solids: {AFLOW}-{AAPL} {Au}tomatic {A}nharmonic {P}honon {Li}brary}, npj\
  Comput.\ Mater. \textbf{3}, 45 (2017).
\input{curtarolo:art125}

\bibitem{Mehl_PRB_91_184100_2015}
M.~J. Mehl, D.~Finkenstadt, C.~Dane, G.~L.~W. Hart, and S.~Curtarolo,
  \emph{Finding the stable structures of N$_{1-x}$W$_x$ with an {\em ab initio}
  high-throughput approach}, Physical Review B \textbf{91}, 184100 (2015).
\input{Mehl_PRB_91_184100_2015}

\bibitem{Bengone_PRB_62_16392_2000}
O.~Bengone, M.~Alouani, P.~Bl\"{o}chl, and J.~Hugel, \emph{Implementation of
  the projector augmented-wave LDA+U method: Application to the electronic
  structure of NiO}, Physical Review B \textbf{62}, 16392 (2000).
\input{Bengone_PRB_62_16392_2000}

\bibitem{Rohrbach_JPCM_15_979_2003}
A.~Rohrbach, J.~Hafner, and G.~Kresse, \emph{Electronic correlation effects in
  transition-metal sulfides}, Journal of Physics: Condensed Matter \textbf{15},
  979--996 (2003).
\input{Rohrbach_JPCM_15_979_2003}

\bibitem{Blochl_PRB_50_17953_1994}
P.~E. Bl\"{o}chl, \emph{Projector augmented-wave method}, Physical Review B
  \textbf{50}, 17953--17979 (1994).
\input{Blochl_PRB_50_17953_1994}

\bibitem{Kresse_PRB_59_1758_1999}
G.~Kresse and D.~Joubert, \emph{From ultrasoft pseudopotentials to the
  projector augmented-wave method}, Physical Review B \textbf{59}, 1758--1725
  (1999). 10.1103/PhysRevB.59.1758.
\input{Kresse_PRB_59_1758_1999}

\bibitem{Perdew_PRL_77_3865_1996}
J.~P. Perdew, K.~Burke, and M.~Ernzerhof, \emph{Generalized Gradient
  Approximation Made Simple}, Physical Review Letters \textbf{77}, 3865--3868
  (1996).
\input{Perdew_PRL_77_3865_1996}

\bibitem{Birch_JGR_83_1257_1978}
F.~Birch, \emph{Finite strain isotherm and velocities for single-crystal and
  polycrystalline nacl at high-pressures and 300-degree-k}, Journal of
  Geophysical Research \textbf{83}, 1257--1268 (1978).
\input{Birch_JGR_83_1257_1978}

\bibitem{Mehl_PRB_47_2493_1993}
M.~J. Mehl, \emph{Pressure dependence of the elastic moduli in aluminum-rich
  {A}l-{L}i compounds}, Physical Review B \textbf{47}, 2493--2500 (1993).
\input{Mehl_PRB_47_2493_1993}

\bibitem{hanson:jmol}
R.~M. Hanson, J.~Prilusky, Z.~Renjian, T.~Nakane, and J.~L. Sussman,
  \emph{Jmol}. An open-source Java viewer for chemical structures in 3D.
\input{hanson:jmol}

\bibitem{Stokes:FROZSL}
H.~T. Stokes, D.~M. Hatch, and i.~B.~J.~Campbell, \emph{FROZSL}. ISOTROPY
  Software Suite.
\input{Stokes:FROZSL}

\bibitem{Mitchell_Engauge_2019}
M.~Mitchell, \emph{Engauge Digitizer} (2019). Open source software, version 12.
\input{Mitchell_Engauge_2019}

\bibitem{Ewald_et_al_Strukturbericht}
P.~P. Ewald, C.~Hermann, O.~Lohrmann, H.~Philipp, C.~Gottfried,
  F.~Schossberger, and K.~Herrmann, eds., \emph{Strukturbericht}, vol. I-VII
  (Akademische Verlagsgesellschaft M. B. H., 1937-1943).
\input{Ewald_et_al_Strukturbericht}

\bibitem{Smithells_Metals_II_1955}
C.~J. Smithells, \emph{Metals Reference Book} (Butterworths Scientific, London,
  1955), second edn.
\input{Smithells_Metals_II_1955}

\bibitem{Yoshiasa_Japanese_J_App_Phys_42_1694_2003}
A.~Yoshiasa, Y.~Murai, O.~Ohtaka, and T.~Katsura, \emph{Detailed Structures of
  Hexagonal Diamond (lonsdaleite) and Wurtzite-type BN}, Japanese Journal of
  Applied Physics \textbf{42}, 1694--1704 (2003).
\input{Yoshiasa_Japanese_J_App_Phys_42_1694_2003}

\bibitem{Hicks_CMS_161_S1_2019}
D.~Hicks, M.~J. Mehl, E.~Gossett, C.~Toher, O.~Levy, R.~M. Hanson, G.~Hart, and
  S.~Curtarolo, \emph{The AFLOW Library of Crystallographic Prototypes: Part
  2}, Computational Materials Science \textbf{161}, S1--S1011 (2019).
\input{Hicks_CMS_161_S1_2019}

\bibitem{Needs_PRB_30_5390_1984}
R.~J. Needs and R.~M. Martin, \emph{Transition from $\beta$-Sn to simple
  hexagonal silicon under pressure}, Physical Review B \textbf{30}, 5390--5392
  (1984).
\input{Needs_PRB_30_5390_1984}

\bibitem{Raynor_Acta_Met_2_616_1954}
G.~V. Raynor and J.~A. Lee, \emph{The tin-rich intermediate phases in the
  alloys of tin with cadmium, indium and mercury}, Acta Metallurgica
  \textbf{2}, 616--620 (1954).
\input{Raynor_Acta_Met_2_616_1954}

\bibitem{Ackland_RPP_64_483_2001}
G.~J. Ackland, \emph{High-pressure phases of group IV and III-V
  semiconductors}, Reports on Progress in Physics \textbf{64}, 483--516 (2001).
\input{Ackland_RPP_64_483_2001}

\bibitem{Wehinger_JPCM_26_115401_2014}
B.~Wehinger, A.~Bosak, G.~Piccolboni, K.~Refson, D.~Chernyshov, A.~Ivanov,
  A.~Rumiantsev, and M.~Krisch, \emph{Diffuse scattering in metallic tin
  polymorphs}, Journal of Physics: Condensed Matter \textbf{26}, 115401 (2014).
\input{Wehinger_JPCM_26_115401_2014}

\bibitem{Hormann_APL_107_123101_2015}
N.~G. H\"{o}rmann, A.~Gross, J.~Rohrer, and P.~Kaghazchi, \emph{Stabilization
  of the $\gamma$-Sn phase in tin nanoparticles and nanowires}, Applied Physics
  Letters \textbf{107}, 123101 (2015).
\input{Hormann_APL_107_123101_2015}

\bibitem{Hedin_J_Phys_C_4_2064_1971}
L.~Hedin and B.~I. Lundqvist, \emph{Explicit local exchange-correlation
  potentials}, Journal of Physics C: Solid State Physics \textbf{4}, 2064--2083
  (1971).
\input{Hedin_J_Phys_C_4_2064_1971}

\bibitem{Ceperley_PRL_45_566_1980}
D.~M. Ceperley and B.~J. Alder, \emph{Ground State of the Electron Gas by a
  Stochastic Method}, Physical Review Letters \textbf{45}, 566--569 (1980).
\input{Ceperley_PRL_45_566_1980}

\bibitem{Perdew_PRB_23_5048_1981}
J.~P. Perdew and A.~Zunger, \emph{Self-interaction correction to
  density-functional approximations for many-electron systems}, Physical Review
  B \textbf{23}, 5048--5079 (1981).
\input{Perdew_PRB_23_5048_1981}

\bibitem{Kohn_PR_140_A1133_1965}
W.~Kohn and L.~J. Sham, \emph{Self-Consistent Equations Including Exchange and
  Correlation Effects}, Physical Review \textbf{140}, A1133--A1138 (1965).
\input{Kohn_PR_140_A1133_1965}

\bibitem{Perdew_PRL_100_136406_2008}
J.~P. Perdew, A.~Ruzsinszky, G.~I. Csonka, O.~A. Vydrov, G.~E. Scuseria, L.~A.
  Constantin, X.~Zhou, and K.~Burke, \emph{Restoring the Density-Gradient
  Expansion for Exchange in Solids and Surfaces}, Physical Review Letters
  \textbf{100}, 136406 (2008).
\input{Perdew_PRL_100_136406_2008}

\bibitem{Armiento_PRB_72_085108_2005}
R.~Armiento and A.~E. Mattsson, \emph{Functional designed to include surface
  effects in self-consistent density functional theory}, Physical Review B
  \textbf{72}, 085108 (2005).
\input{Armiento_PRB_72_085108_2005}

\bibitem{Mattsson_JCP_128_084714_2008}
A.~E. Mattsson, R.~Armiento, J.~Paier, G.~Kresse, J.~M. Wills, and T.~R.
  Mattsson, \emph{The AM05 density functional applied to solids}, Journal of
  Chemical Physics \textbf{128}, 084714 (2008).
\input{Mattsson_JCP_128_084714_2008}

\bibitem{Tao_PRL_91_146401_2003}
J.~Tao, J.~P. Perdew, V.~N. Staroverov, and G.~E. Scuseria, \emph{Climbing the
  Density Functional Ladder: Nonempirical Meta–Generalized Gradient
  Approximation Designed for Molecules and Solids}, Physical Review Letters
  \textbf{91}, 146401 (2003).
\input{Tao_PRL_91_146401_2003}

\bibitem{Perdew_PRL_103_206403_2009}
J.~P. Perdew, A.~Ruzsinszky, G.~I. Csonka, L.~A. Constantin, and J.~Sun,
  \emph{Workhorse Semilocal Density Functional for Condensed Matter Physics and
  Quantum Chemistry}, Physical Review Letters \textbf{103}, 026403 (2009).
\input{Perdew_PRL_103_206403_2009}

\bibitem{Sun_JCP_137_051101_2012}
J.~Sun, B.~Xiao, and A.~Ruzsinszky, \emph{Communication: Effect of the
  orbital-overlap dependence in the meta generalized gradient approximation},
  Journal of Chemical Physics \textbf{137}, 051101 (2012).
\input{Sun_JCP_137_051101_2012}

\bibitem{Sun_JCP_138_044113_2013}
J.~Sun, R.~Haunschild, B.~Xiao, I.~W. Bulik, G.~E. Scuseria, and J.~P. Perdew,
  \emph{Semilocal and hybrid meta-generalized gradient approximations based on
  the understanding of the kinetic-energy-density dependence}, Journal of
  Chemical Physics \textbf{138}, 044113 (2013).
\input{Sun_JCP_138_044113_2013}

\bibitem{Zhao_JCP_125_194101_2006}
Y.~Zhao and D.~G. Truhlar, \emph{A new local density functional for main-group
  thermochemistry, transition metal bonding, thermochemical kinetics, and
  noncovalent interactions}, Journal of Chemical Physics \textbf{125}, 194101
  (2006).
\input{Zhao_JCP_125_194101_2006}

\bibitem{Sun_PRL_115_036402_2015}
J.~Sun, A.~Ruzsinszky, and J.~Perdew, \emph{Strongly Constrained and
  Appropriately Normed Semilocal Density Functional}, Physical Review Letters
  \textbf{115}, 036402 (2015).
\input{Sun_PRL_115_036402_2015}

\bibitem{Sun_Nature_Chem_8_831_2016}
J.~Sun, R.~C. Remsing, Y.~Zhang, Z.~Sun, A.~Ruzsinszky, H.~Peng, Z.~Yang,
  A.~Paul, U.~Waghmare, X.~Wu, M.~L. Klein, and J.~P. Perdew, \emph{Accurate
  first-principles structures and energies of diversely bonded systems from an
  efficient density functional}, Nature Chemistry \textbf{8}, 831--836 (2016).
\input{Sun_Nature_Chem_8_831_2016}

\bibitem{Csonka_PRB_79_155107_2009}
G.~I. Csonka, J.~P. Perdew, A.~Ruzsinszky, P.~H.~T. Philipsen, S.~Leb\`{e}gue,
  J.~Paier, O.~A. Vydrov, and J.~G. \'{A}ngy\'{a}n, \emph{Assessing the
  performance of recent density functionals for bulk solids}, Physical Review B
  \textbf{79}, 155107 (2009).
\input{Csonka_PRB_79_155107_2009}

\bibitem{Sun_PRB_84_035117_2011}
J.~Sun, M.~Marsman, G.~I. Csonka, A.~Ruzsinszky, P.~Hao, Y.-S. Kim, G.~Kresse,
  and J.~P. Perdew, \emph{Self-consistent meta-generalized gradient
  approximation within the projector-augmented-wave method}, Physical Review B
  \textbf{84}, 035117 (2011).
\input{Sun_PRB_84_035117_2011}

\bibitem{Deshpande_Acta_Cryst_14_355_1961}
V.~T. Deshpande and D.~B. Sirdeshmukh, \emph{Thermal Expansion of Tetragonal
  Tin}, Acta Crystallographica \textbf{14}, 355--356 (1961).
\input{Deshpande_Acta_Cryst_14_355_1961}

\bibitem{Price_PRB_3_1268_1971}
D.~L. Price, J.~M. Rowe, and R.~M. Nicklow, \emph{Lattice Dynamics of Grey Tin
  and Indium Antimonide}, Physical Review B \textbf{3}, 1268--1279 (1971).
\input{Price_PRB_3_1268_1971}

\bibitem{Christensen_PRB_48_5797_1993}
N.~E. Christensen and M.~Methfessel, \emph{Density-functional calculations of
  the structural properties of tin under pressure}, Physical Review B
  \textbf{48}, 5797--5807 (1993).
\input{Christensen_PRB_48_5797_1993}

\bibitem{Olijnyk_J_Phys_Coll_45_C8_153_1984}
H.~Olijnyk and W.~B. Holzapfel, \emph{Phase Transitions in Si, Ge and Sn Under
  Pressure}, Suppl\'{e}ment au Journal de Physique Colloques \textbf{45},
  153--156 (1984).
\input{Olijnyk_J_Phys_Coll_45_C8_153_1984}

\bibitem{Rowe_PRL_14_554_1965}
J.~M. Rowe, B.~N. Brockhouse, and E.~C. Svensson, \emph{Lattice Dynamics of
  White Tin}, Physical Review Letters \textbf{14}, 554--556 (1965).
\input{Rowe_PRL_14_554_1965}

\bibitem{Price_Proc_Roy_Soc_Lon_A_300_25_1967}
D.~L. Price, \emph{Lattice Dynamics of White Tin}, Proceedings of the Royal
  Society of London. Series A, Mathematical and Physical Sciences \textbf{300},
  25--44 (1967).
\input{Price_Proc_Roy_Soc_Lon_A_300_25_1967}

\bibitem{Setyawan_CMS_49_299_2010}
W.~Setyawan and S.~Curtarolo, \emph{High-throughput electronic band structure
  calculations: Challenges and tools}, Computational Materials Science
  \textbf{49}, 299--312 (2010).
\input{Setyawan_CMS_49_299_2010}

\bibitem{Touloukian_Thermal_Properties_1975}
Y.~S. Touloukian, R.~K. Kirby, R.~E. Taylor, and P.~D. Desai, \emph{Thermal
  {Expansion} {Metallic} {Elements} and {Alloys}} (New York : IFI/Plenum,
  1975), \emph{Thermophysical {Properties} of {Matter} - the {TPRC} {Data}
  {Series}}, vol.~12, pp. 339--345.
\input{Touloukian_Thermal_Properties_1975}

\bibitem{Bachelet_PRB_26_4199_1982}
G.~B. Bachelet, D.~R. Hamann, and M.~Schl\"{u}ter, \emph{Pseudopotentials that
  work: From H to Pu}, Physical Review B \textbf{26}, 4199--4228 (1982).
\input{Bachelet_PRB_26_4199_1982}

\bibitem{Liechtenstein_PRB_52_R5467_1995}
A.~I. Liechtenstein, V.~I. Anisimov, and J.~Zaanen, \emph{Density-functional
  theory and strong interactions: Orbital ordering in Mott-Hubbard insulators},
  Physical Review B \textbf{52}, R5467--R5470 (1995).
\input{Liechtenstein_PRB_52_R5467_1995}

\bibitem{Calderon_Comp_Mat_Sci_108A_233_2015}
C.~E. Calderon, J.~J. Plata, C.~Toher, C.~Oses, O.~Levy, M.~Fornari, A.~Natan,
  M.~J. Mehl, G.~Hart, M.~B. Nardelli, and S.~Curtarolo, \emph{The \{AFLOW\}
  standard for high-throughput materials science calculations}, Computational
  Materials Science \textbf{108, Part A}, 233--238 (2015).
\input{Calderon_Comp_Mat_Sci_108A_233_2015}

\end{thebibliography}
\end{document}